\documentclass[aps,showpacs,twocolumn,superscriptaddress,prb,floatfix, 10pt]{revtex4-1}
\usepackage{graphicx}
\usepackage{subfigure}
\usepackage{amssymb,amsmath,cancel}
\usepackage{comment}
\usepackage{xcolor}
\usepackage{array}

\begin{document}

\title{Instanton confinement-deconfinement transitions: The stability of pseudogap phases and topological order}

\author{Predrag Nikoli\'c}
\affiliation{Department of Physics and Astronomy,\\George Mason University, Fairfax, VA 22030, USA}
\affiliation{Institute for Quantum Matter at Johns Hopkins University, Baltimore, MD 21218, USA}

\date{\today}

%
%
%

\begin{abstract}

We explore the stability of certain many-body quantum states which may exist at zero or finite temperatures, may lack long-range order and even topological order, and still are thermodynamically distinct from uncorrelated disordered phases. We sharply characterize such states by the conservation of topological charge, or equivalently confinement of instantons, using a generalization of the Wilson loop and the correlation length of an emergent gauge field. Our main conclusions are: (i) topological orders can exist at finite temperatures, (ii) relativistic liquids of topological defects can also exist as stable phases at finite temperatures, and (iii) there are two universality classes of instanton suppression. We also relate the instanton dynamics to the problem of the pseudogap state in underdoped cuprates. A universal experimental signature of the instanton deconfinement transition is a change of the quantum noise spectrum, which can perhaps be measured in some situations, for example via a quantum anomaly, or indirectly detected with a specific heat jump. The method of analysis is a functional renormalization group that generalizes the Coulomb gas treatment of Kosterlitz and Thouless to arbitrary interactions and dimensions. In particular, we construct an exact non-perturbative technique for confining interactions between instantons, which introduce irreparable infra-red divergences in the standard perturbative approaches.

\end{abstract}

\maketitle

\section{Introduction}

Strongly correlated states of matter pose a great challenge to our universal understanding of materials, especially when interactions between particles are combined with quantum entanglement, thermal effects and non-trivial topology. Topological order is a form of many-body quantum entanglement which cannot be unambiguously detected by correlations between any local properties, and yet produces ground state degeneracy on topologically non-trivial manifolds \cite{WenQFT2004}. Fractional quantum Hall states are the only experimentally confirmed realizations of topological order in electronic materials up to date. However, the theory of topological orders has advanced very far. A complete classification of topological orders in two spatial dimensions may have been achieved \cite{Wen2017}, and at least a partial classification in three dimensions is emerging \cite{Levin2009, Maciejko2010, Swingle2011, Walker2012, Keyserlingk2013, Jian2014, Wang2014a, Ye2015, Ye2016, Putrov2017, Furusaki2019, Wen2019a, Wen2019}. This major gap between theory and experiment is actually also manifest in the theory itself since it seems unclear how to take a random case from the available classification schemes and construct a realistic microscopic model (beyond a specialized exactly solvable model) which realizes the chosen topological order in its ground state. This is in contrast to the phenomenon of symmetry breaking and conventional long-range order: the moment we choose a group for symmetry transformations, we have an idea about what degrees of freedom to use and how to couple them within a physical Hamiltonian, and then we have many techniques at our disposal to analyze the dynamics. The present study attempts to approach the problem of topological order from a similar physical point of view, still quite universal but focused on the issues of feasibility and stability instead of classification.

Here, we explore certain general conditions for the stability of unconventional quantum phases which exhibit strong correlations and no spontaneous symmetry breaking. We are interested in the phases shaped by delocalized but uncondensed topological defects. A topological protection mechanism that conserves topological charge, and hence allows particles to be delocalized as well, is engaged through the confinement of instantons. We study instanton confinement as a function of temperature and interaction type in $D=d+1>2$ space-time dimensions using renormalization group. While much is already known about the issues of confinement \cite{Berezinsky1972, Kosterlitz1973, Kosterlitz1974, Wilson1974, Polyakov1975, Belavin1975, tHooft1976, Kosterlitz1977, Polyakov1977, Polyakov1978, Osterwalder1978, Guth1980, Lautrup1980, Yaffe1982, Nagaosa1993, Shuryak1998, Kondo1998, zheng89, read90, Kondo1999, tHooft2000, Nagaosa2000, Yoshida2001, sachdev02e, Sudbo2002, Sudbo2004, Herbut2003, Herbut2004, herbut03, Hermele2004, Sudbo2005, Engelhardt2005, Wang2005a, Kleinert2005, Nogueira2008, Yamamoto2008, Oglivie2012, Anber2014, Heydeman2023}, the general scope and various technical aspects of this study make it non-trivial. The instanton context of confinement leads to several important and surprising conclusions: (i) topological order is possible at finite temperatures, (ii) relativistic correlated quantum liquids of topological defects and anti-defects are also possible at finite temperatures, with or without topological order, and (iii) there is more than one universality class of instanton deconfinement.

We will show that stable thermodynamic phases of confined instantons can exist independently of spontaneous symmetry breaking in generic systems which support topological defects. An important prerequisite is the presence of an intrinsic length scale distinct from the ultra-violet cut-off and unrelated to temperature or disorder. The new length scale is required in order to determine a finite coherence length $\xi$ of the matter field. If the instantons are neutralized within distances $\lambda<\xi$, then topological defects are well-defined and conserved for all practical purposes. The fractional quantum Hall state is an example: it lacks long-range order at least because the introduced vortices are mobile, but its filling factor $\nu$, the number of particles per vortex, is rationally quantized. Since charge is conserved, this is possible only if topological charge, or vorticity, is conserved as well. Another example may be the Nernst effect in the underdoped pseudogap state of cuprates \cite{Ong2001, Wang2006, Li2010}, which demonstrates coherent vortex dynamics in the absence of a Cooper pair condensate. Our findings give a theoretical support for such correlation phenomena, generalizing the thermodynamics of protected topological defects to finite temperatures and beyond topological order. We suggest that an instanton confinement phase transitions can be experimentally detected with quantum noise and specific heat measurements, at least in principle.

This paper is organized as follows. Section \ref{secTDyn} introduces instantons and explains a thermodynamic characterization of their confinement with a generalized Wilson operator and a correlation length for an emergent gauge field. Then, Section \ref{secCases} surveys the literature about confinement-deconfinement transitions and describes the context of this study together with its essential ideas for the setup and interpretation of renormalization group. Technical presentation of the instanton gas renormalization group is given in Section \ref{secRG} (which uninterested readers can skip). After a brief introduction, we consider non-confining potentials in Section \ref{secNonConf}, logarithmic potentials in Section \ref{secLog}, general confining potentials in Section \ref{secConf}, and finite-temperature phase transitions in Section \ref{secFiniteTemp}. At the end, the critical scaling of the correlation length with temperature is derived for two universality classes of instanton deconfinement. Then, Section \ref{secExp} explores possible experimental signatures of the confined-instanton phases, hidden in the quantum noise. Section \ref{secDiscussion} is a non-technical discussion of possible implications for the physical systems such as fermionic and bosonic quantum Hall liquids, high-temperature superconductors, and materials with strong spin-orbit coupling. All conclusions are summarized in Section \ref{secConclusions}.

Throughout this paper, we use units $\hbar=c=1$ and Einstein's convention for the summation over repeated indices. Spatial directions are denoted by Latin letters $i,j,k,\dots \in \lbrace 1, \dots, d \rbrace$, and space-time directions by Greek letters $\mu,\nu,\lambda,\cdots\in \lbrace 0, 1, \dots, d \rbrace$.

\section{Thermodynamic distinction between confined and deconfined phases}\label{secTDyn}

Consider a generic quantum system in the continuum limit whose local degrees of freedom are accessed by a field operator $\psi({\bf x})$. If the system supports topological defects, it is also possible to construct an operator $\mathcal{J}_0\lbrack\psi\rbrack$ which probes their  density. The topological index or ``charge'' of all defects inside a $d$-dimensional volume $B^d$ is
\begin{equation}
N = \int\limits_{B^d} d^dx\,\mathcal{J}_0 \ .
\end{equation}
No smooth field deformations can change the topological index. We mathematically represent this fact by a conservation law $\partial_\mu \mathcal{J}_\mu=0$, where $\mathcal{J}_i$ is the current density of topological defects. The solution of the conservation equation can be expressed using an antisymmetric tensor gauge field $\mathcal{A}_{\mu_1\cdots\mu_{d-1}}$,
\begin{equation}\label{SingularityGauge}
\mathcal{J}_\mu = \epsilon_{\mu\nu\lambda_1\cdots\lambda_{d-1}} \partial_\nu \mathcal{A}_{\lambda_1\cdots\lambda_{d-1}} \ .
\end{equation}
Explicit constructions of $\mathcal{J}_\mu$ have been obtained for spinor and vector fields whose topological defects (vortices, monopoles and hedgehogs) are classified by the $\pi_n(S^n)$ homotopy groups \cite{Nikolic2019}. Using Stokes-Cartan theorem, we find that the topological invariant is the gauge flux through the boundary $S^{d-1}$ of $B^d$:
\begin{eqnarray}
N &=& \int\limits_{B^d} d^dx\,\epsilon_{0ij_1\cdots j_{d-1}} \partial_i \mathcal{A}_{j_1\cdots j_{d-1}} \\
&=& \oint\limits_{S^{d-1}} d^{d-1}x\,\epsilon_{j_1\cdots j_{d-1}} \mathcal{A}_{j_1\cdots j_{d-1}} \ . \nonumber
\end{eqnarray}
In our notation, the Levi-Civita tensor $\epsilon$ either carries all space-time indices, or displays only the indices which live on the integration manifold as in the last line.

In order to study dynamics at any temperature, we construct the imaginary-time path integral for the partition function
\begin{equation}\label{PartFunct}
\mathcal{Z} = \textrm{tr}(e^{-\beta H\lbrack\psi\rbrack}) = \int \mathcal{D}\psi\,e^{-S\lbrack\psi\rbrack} \ .
\end{equation}
Here, $H$ is the Hamiltonian operator, $S$ is the imaginary-time action, and $\beta=1/T$ is the inverse temperature. The action is an integral of the Lagrangian density $\mathcal{L}$ over the infinite $d$-dimensional space and a finite imaginary time extent $\tau\in(0,\beta)$ with periodic boundary conditions. Let us introduce the instanton density operator
\begin{equation}
\mathcal{I} = \partial_\mu \mathcal{J}_\mu \ ,
\end{equation}
which vanishes if the topological charge is conserved. For our purposes, instantons are the lowest-action events of quantum tunneling between topologically distinct classical configurations, i.e. events in which topological defects are created or annihilated. Therefore, instantons require singularities in the field configuration $\psi(x_\mu)$. We can apply Gauss' theorem to a space-time volume $B^{d+1}$ and express the number of instantons in it as the flux of $\mathcal{J}_\mu$ through the boundary $S^{d}$:
\begin{equation}\label{InstantonCharge}
\mathcal{N} = \int\limits_{B^{d+1}} d^{d+1}x\, \mathcal{I} = \oint\limits_{S^d} d^dx\, \hat{\eta}_\mu \mathcal{J}_\mu \ ,
\end{equation}
where $\hat{\eta}_\mu$ is the unit-vector locally perpendicular to the oriented surface $S^d$.

By simple symmetry considerations, we could naively capture the essential dynamics of topological defects with a gradient term $\mathcal{J}_\mu\mathcal{J}_\mu$ in the Lagrangian density, which is the Maxwell term of the gauge field $\mathcal{A}_{\mu_1\cdots\mu_{d-1}}$. If no other Lagrangian term is relevant for instanton dynamics, then the action cost of $\mathcal{N}\neq 0$ will be minimized by evenly distributing the flux of $\mathcal{J}_\mu$ over the entire $S^d$ manifold. This produces a Coulomb attractive interaction potential between opposite-charge instantons separated by $r$ in the $D=d+1$ dimensional space-time:
\begin{equation}
V_{\textrm{C}}(r) \sim 1/r^{D-2}
\end{equation}
This estimate comes from the $\mathcal{N}=1$ solution of (\ref{InstantonCharge}), $\mathcal{J}_\mu(r) \propto \hat{r}_\mu / r^d$, and the ensuing action cost $\int d^{d+1}x\, (\mathcal{J}_\mu)^2 \propto 1/r^{d-1}$ cut-off by the distance between the two instantons. It is known that the Coulomb interaction in $D>2$ is not able to confine its charged particle sources into neutral dipoles \cite{Kosterlitz1977}. Consequently, instantons with this interaction would be free to roam and destroy the conservation of topological charge at macroscopic length and time scales.

\begin{figure}
\subfigure[{}]{\includegraphics[width=1.6in]{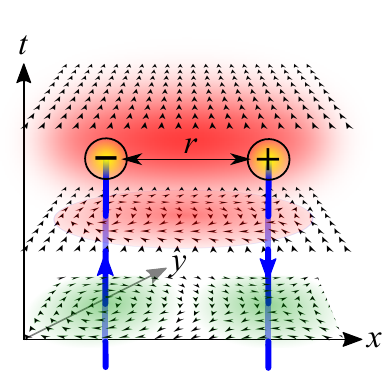}}\hspace{0.1in}
\subfigure[{}]{\includegraphics[width=1.6in]{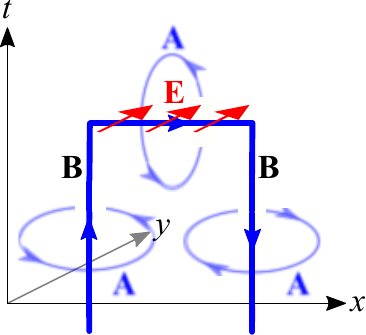}}
\caption{\label{FigInst}An illustration of instantons for vortices in $D=2+1$ space-time dimensions. (a) Worldlines of a vortex and antivortex (blue) are terminated by a pair of opposite-charge instantons. In this example, the two instantons occur at the same time $t=t_0$, but this is not generally required. The phase gradients $\boldsymbol{\nabla}\theta$ of $\psi\sim e^{i\theta}$, which build the action cost, are indicated with colored clouds. Far away from the instantons, the phase gradients are minimal (green), and typically produce a Coulomb-like interaction between the topological defects (screened to a finite range in the presence of gauge fields). But, the gradients must smoothly deform near the instantons in order to minimize the impact of a sudden change across $t=t_0$. In the best case scenario, there is a large action cost $V(r)$ on the temporal space-time lattice links (due to $\partial_t\theta$) at the time $t=t_0$ and within a region of space which grows in proportion to the distance $r$ between the instantons. The resulting $V(r)\propto r$ confines the instantons and suppresses the shown vortex creation/annihilation events. However, the underlying assumption was that discontinuous changes of $\theta$ are very costly. Instead, phase fluctuations on a lattice can be abundant enough to uncorrelate $\theta$ even across distances comparable to the lattice constant. This situation characterizes Mott insulators and removes the linear confinement of instantons (i.e., $V(r)$ becomes short-ranged). (b) The case of confined instantons: vortex flux lines cannot terminate. In this example, a vortex-antivortex pair annihilates, but the horizontal connecting flux line $\boldsymbol{\nabla}\times{\bf A}\neq 0$ corresponds to the electric field ${\bf E}=\partial_t{\bf A}$ which is generated via Faraday's law. The Maxwell action cost of this electric field suppresses vortex creation/annihilation process across large distances.}
\end{figure}

However, the above picture is inconsistent with the smooth low-energy configurations of the matter field $\psi$; it requires a continuous distribution of singularities. Instead, a coherent $\psi$ field, governed by a $(\partial_\mu \psi)^2$ Lagrangian density, focuses its non-zero topological currents into lower-dimensional singular domains. The flux of $\mathcal{J}_\mu$ satisfying (\ref{InstantonCharge}) then stretches through a microscopically narrow tube, as a string connecting the two opposite-charge instantons. The string acquires tension from the depletion of the coherent $\psi$ background inside the tube, leading to a linear interaction between instantons:
\begin{equation}
V_{\textrm{conf.}}(r) \sim r \ .
\end{equation}
An example with vortices is shown in Fig.\ref{FigInst}. This kind of interaction is confining. Instantons of opposite charge are asymptotically free at short distances but cannot drift away from each other. Their dynamics is, however,  complicated by the instanton dipole fluctuations. If one tries to separate a pair of opposite-charge test instantons to an arbitrarily large distance, one eventually supplies enough action to generate another dipole of instantons. The generated instantons can move to completely neutralize the original instantons, and allow one to continue pulling them apart without further resistance. Therefore, screening is inevitable despite a finite action cost or ``mass'' of an instanton. The resulting finite screening length $\lambda$ may be further reduced by virtual instanton fluctuations.

We will show that instanton confinement is a thermodynamic phase transition which can occur at finite temperatures independently of symmetry breaking. A sharp distinction between the confined and deconfined phases is given by a generalization of the Wilson loop \cite{Wilson1974} operator defined on the closed space-time surface $S^d$. This operator counts the total instanton charge enclosed by $S^d$, and we can set it up in imaginary time using (\ref{InstantonCharge}) at zero temperature:
\begin{equation}\label{Wilson}
\mathcal{C}(S^d) \equiv \mathcal{N} \ .
\end{equation}
This is a random variable and we are interested in its variance. If instantons are deconfined, then there is a finite probability of an uncompensated instanton appearing at any point in space-time, uncorrelated with the locations of other instantons. Such random fluctuations make the variance $\textrm{Var}\,\mathcal{C}(S^d) \propto B^{d+1}$ scale as the volume $B^{d+1}$ of the space-time enclosed by $S^d$. In contrast, confined instantons are always compensated within a finite radius $\lambda$, so their number inside $S^d$ can be affected only by small dipole fluctuations near $S^d$, giving rise to an ``area-law'' behavior $\textrm{Var}\,\mathcal{C}(S^d) \propto S^d$. This distinction is made in the limit of infinite surface $S^d\to\infty$, so it is impossible to smoothly deform the ``volume-law'' into the ``area-law'' behavior. Ultimately, $\textrm{Var}\,\mathcal{C}(S^d)$ serves the role of an ``order parameter'' for the thermodynamic phase transition between the confined and deconfined states of instantons:
\begin{equation}\label{Wilson2}
\textrm{Var}\,\mathcal{C}(S^d) \xrightarrow{S^d\to\infty}{}
  \begin{cases}
    a S^d &,\quad \textrm{confined phase} \\
    b B^{d+1} & ,\quad \textrm{deconfined phase}
\end{cases} \ .
\end{equation}

The most general definition of the operator $\mathcal{C}(S^d)$ is obtained in real time by exploiting the formal correspondence between the imaginary and real time path integrals at zero temperature. We just interpret (\ref{InstantonCharge}) as a quantum mechanical operator expressed in terms of the field (creation/annihilation) operators $\psi$, and replace the imaginary time coordinate $x_0\equiv\tau$ with real time $x_0\to it$:
\begin{equation}\label{Wilson3}
\mathcal{C}(S^d) = \oint\limits_{S^d} d^dx\, \hat{n}_\mu \mathcal{J}_\mu\lbrack\psi\rbrack \ .
\end{equation}
Note that a factor of $i$ inherited from the integral measure was absorbed into the appropriate (re)definition of $\mathcal{J}_\mu$ as a Hermitian operator. The time dependence of field operators is handled as usual with $\psi(t)=e^{iHt}\psi(0)e^{-iHt}$ in the Heisenberg picture. Measurements of $\mathcal{C}(S^d)$ in the ground state $|0\rangle$ yield random outcomes, and their variance follows (\ref{Wilson2}) by the same topological considerations as before. Having a proper quantum operator prescribes measurements at any temperature as
\begin{equation}
\textrm{Var}\,\mathcal{C} \equiv \bigl\langle (\mathcal{C} - \langle\mathcal{C}\rangle)^2 \bigr\rangle
\quad,\quad \langle \cdots \rangle = \textrm{tr} (\cdots e^{-\beta H}) \ ,
\end{equation}
even when the exact real-imaginary time correspondence is lost due to $\beta<\infty$. This enables a sharp distinction between confined and deconfined phases at $T>0$; the infinite extent of real time allows taking the $S^d\to\infty$ limit.

Another important characterization of the topological defect dynamics is the correlation length $\zeta$ of the effective gauge field $\mathcal{A}_{\lambda_1\cdots\lambda_{d-1}}$. When instanton fluctuations are suppressed, i.e. $\partial_\mu \mathcal{J}_\mu = 0$, then the Maxwell Lagrangian density $\frac{1}{2e^{2}}\mathcal{J}_{\mu}\mathcal{J}_{\mu}$ with (\ref{SingularityGauge}) yields a gauge-invariant correlation function in momentum space
\begin{eqnarray}\label{PhotonCorrel}
&& \left\langle \mathcal{A}_{\mu_{1}\cdots\mu_{d-1}}({\bf q})\mathcal{A}_{\nu_{1}\cdots\nu_{d-1}}({\bf q}')\right\rangle \sim \\
&& \qquad \sim \frac{e^{2}}{q^{2}}\frac{q_{\alpha}q_{\beta}}{q^{2}}\epsilon_{\lambda\alpha\mu_{1}\cdots\mu_{d-1}}\epsilon_{\lambda\beta\nu_{1}\cdots\nu_{d-1}}\,\delta({\bf q}+{\bf q}') \nonumber \ .
\end{eqnarray}
This algebraic form signifies the presence of gapless photon modes and a formally infinite correlation length that characterizes the gauge field fluctuations. Note that a topological Lagrangian term which arises from a Berry curvature, or the Higgs mechanism, can open a gap in the photon spectrum, but this may also modify the instanton interaction potential. If uncompensated instantons are present and separated by a mean space-time distance $\zeta$, then the photons can propagate coherently only across the length and time scales of $\zeta$. This effectively gives the photons a mass $m\sim\zeta^{-1}$. We will find that the correlation length $\zeta$ diverges as a power of $T-T_c$ when the temperature $T$ approaches from above its critical value $T_c$ for the instanton confinement transition.

\subsection{Cases of confinement and deconfinement}\label{secCases}

Most studies of confinement are specialized to a particular gauge theory, and ask if the gauge field fluctuations produce a potential energy $U(r)$ between static test charges which grows indefinitely with the distance $r$. The common Abelian electrodynamics ($D=3+1$) does not confine charge, but its compact variety, whose lattice action 
\begin{equation}\label{CompactU1}
S = -\frac{1}{g^2} \sum_\square \cos\left( \sum_{\mu\in \square} A_\mu \right)
\end{equation}
evaluates the gauge flux on lattice plaquettes $\square$, causes confinement in the strong coupling $g\gg 1$ regime \cite{Wilson1974, Polyakov1975, Osterwalder1978, Guth1980, Lautrup1980}. Confinement-deconfinement transitions are found in all $D\ge 4$ compact U(1) gauge theories \cite{Yaffe1982}, while the $D=3$ case is special and always confined in the absence of a dynamical matter field \cite{Polyakov1975}. Here is a naive explanation. The weak coupling regime $g\ll 1$ justifies taking the continuum limit
\begin{equation}
S\to \frac{1}{2\gamma^2} \int d^Dx\, (\epsilon_{\cdots \mu\nu} \partial_\mu A_\nu)^2 \ .
\end{equation}
This non-compact Maxwell form suppresses monopoles, i.e. terminations of the vortex (world)lines, and hence mediates plain Coulomb interactions between charges. It just happens that the Coulomb potential $U(r) \propto 1/r^{d-2}$ between particles separated by $r$ vanishes as $r\to \infty$ in $d\ge 3$, but becomes confining $U(r) \propto \log (r/a)$ in $d=2$ ($D=3$), making only neutral bound states possible. The opposite $g\gg 1$ limit of (\ref{CompactU1}) has to be considered on the lattice. It admits free flux-quantum fluctuations through lattice plaquettes and hence allows monopoles. Abundant monopoles that cannot be removed by a gauge transformation strongly frustrate the dynamics of uncompensated charges via quantum interference, until only neutral objects can exist as low-energy excitations. The mechanism is similar to the particle localization in Mott insulators when viewed from a dual theory \cite{Fisher1989a, Fisher1990}. Since gauge symmetry breaking is not possible \cite{Elitzur1975}, the phase transitions in pure gauge theories have topological character.

The deconfinement of particles in the presence of dynamical matter fields has been intensely studied in condensed matter physics \cite{fradkin79a, Nagaosa2000}, especially in relation to the stability of spin liquids \cite{Sudbo2002, Sudbo2004, Herbut2003, Herbut2004, herbut03, Hermele2004, Nogueira2008}. Massless particles suppress gauge field fluctuations and facilitate charge deconfinement. In $d=2$ spatial dimensions, where this matters most, massless Dirac fermions coupled to a U(1) gauge field can be deconfined if they exist in sufficiently many flavors \cite{Hermele2004, Nogueira2008}. Fermi liquids expel electric fields and superconductors expel all gauge fields, so their particles are self-consistently deconfined. Everything becomes more complicated with non-Abelian gauge fields. A crucial property of Yang-Mills gauge theory is the existence of instanton configurations which cost a finite action even in the continuum limit \cite{Belavin1975, tHooft1976, Shuryak1998, tHooft2000, Oglivie2012}. Similar to skyrmions in $d=2$, these instantons are characterized by an arbitrary length-scale which determines their size, so they are relevant macroscopically (in the infra-red limit). Their proliferation is responsible for the confinement of quarks.

Our interest in this paper is the confinement of instantons instead of particles. The fate of instantons is most directly decided by their interaction potential. We only tackle a small part of this problem by asking which instanton interactions lead to confinement, and what kinds of deconfinement transitions occur in $D$ space-time dimensions at zero or finite temperatures. We answer these questions by a real-space ``instanton gas renormalization group'' (IGRG), building on the Coulomb gas renormalization group of Kosterlitz and Thouless \cite{Kosterlitz1973, Kosterlitz1974, Kosterlitz1977}. In contrast to many similar studies in soft condensed matter contexts \cite{Levesque1986, Minnhagen1987, Fisher1994, Fisher1994b, Nussinov2023}, we take into account the dipole creation/annihilation processes.

Most modern works address the charge or monopole confinement by calculating the renormalization of a bare interaction potential without considering the possibility of losing the infra-red convergence of the partition function. This might not be a problem for the typical studied cases of non-confining potentials and logarithmic potentials in no more than three dimensions \cite{Sudbo2002, Sudbo2004, Herbut2003, Herbut2004, Heydeman2023}. However, it definitely becomes a problem when perturbative renormalization group arguments are generalized to confining potentials. The main technical distinction of this work from prior literature is the systematic exposure of infra-red divergences in Sections \ref{secNonConf} and \ref{secLog}, and the proposal of a remedy for confining potentials in Section \ref{secConf}.

Instanton interaction potentials which vanish at large distances are regarded as naively non-confining. As expected \cite{Kosterlitz1977}, we find that the Coulomb and weaker potentials cannot confine instantons in any $D$ dimensions at any temperature. The employed perturbative renormalization group method breaks down for stronger non-confining potentials due to the emergence of couplings which introduce an infra-red divergence. This could indicate renormalization into confining potentials, but must be verified with different approaches.

We show that infra-red divergences have a universal resolution in the emergence of exclusive confining interactions: instantons become connected by ``flux tubes'' and interact with only one partner of opposite charge, like quarks in mesons. This can be analyzed with an exact real-space renormalization group. We find that exclusive potentials which are asymptotically free and more confining than the logarithm generally induce instanton confinement below a finite critical temperature. Exclusive confining potentials which are singular at the origin, such as the logarithm, are considered not asymptotically free. They stabilize a different fixed point, with deconfined but suppressed instantons at zero temperature and no finite temperature transition.

Spontaneous symmetry breaking generally gaps out gauge fields and thus suppresses the instantons. As explained earlier, the interaction between instantons is then confining, in fact linear $V(r)\propto r$ if the instantons create and destroy singular point defects with a $\pi_{d-1}(S^{d-1})$ homotopy invariant (see Appendix \ref{app1} for a general derivation). Instantons associated with the Hopf index interact at least logarithmically, which is still confining (shown in Appendix \ref{app2}), but instantons associated with finite loops and other contractible structures tend to have short-range interactions. If the potential is confining, every instanton will be compensated by its anti-instanton within a finite screening radius $\lambda$. Even when fluctuations or Pauli exclusion preclude the broken-symmetry phase, the dynamics can retain coherence of the ordered state below the length and time scales given by a correlation length $\xi$. If $\lambda<\xi$ is maintained, we may compute the confining instanton interaction potential as if we were in the ordered phase and then let the renormalization group confirm confinement from the macroscopic point of view by showing the flow $\xi>\lambda \to 0$. We, therefore, find instanton confinement in the absence of spontaneous symmetry breaking, and sharply distinguish it from the opposite regime $\lambda\gg \xi$ by the generalized Wilson correlation (\ref{Wilson2}).

\section{Instanton gas renormalization group}\label{secRG}

In order to study instanton confinement, we introduce instantons at space-time positions ${\bf x}_i$ as boundary conditions on the $\psi$ field in the partition function (\ref{PartFunct}). Then we integrate out the smooth fluctuations of $\psi$ in order to obtain the instanton path integral:
\begin{equation}\label{IPartFunct}
\mathcal{Z} =\sum_{N=0}^{\infty}\frac{y_{0}^{N}}{R_{0}^{DN}}\frac{1}{N_+!N_-!}\int\prod_{i=1}^{N}d^{D}x_{i}\,e^{-S_N({\bf x}_1,\dots,{\bf x}_N)} \ .
\end{equation}
The configurations of $N_\pm$ identical instantons of charge $\pm$ are still summed over, with $N_++N_-=N$. The action $S_0$ of an instanton core is consumed into the fugacity $y_0=e^{-S_0}$, and an ultra-violet cut-off length scale $R_0$ is included in order to keep $y_0$ dimensionless. We will not attempt to derive the effective instanton action $S_N$ from any microscopic model. Instead, we will consider all possible interaction potentials and attempt to classify all possible scenarios that lead to confinement or deconfinement. We will neglect three-body and other many-body interactions with an assumption that interactions over large distances are the ones most important for the macroscopic behavior of the system.

The usual picture of non-exclusive pair-wise interactions
\begin{equation}\label{NonExclusive}
S_N = \frac{1}{2}\sum_{i\neq j}^{1\dots N} q_i q_j V ({\bf x}_i - {\bf x}_j)
\end{equation}
works well for non-confining potentials that vanish at large distances, $V(r\to\infty)\to 0$. Here, $q_i=\pm 1$ is the instanton charge at position ${\bf x}_i$. However, this breaks down for confining potentials $V(r\to\infty)\to \infty$ which allow only bound states: their screening computed from the partition function is plagued by an infra-red divergence. We will show that physically acceptable confining interactions must be exclusive and act only between opposite-charge instantons:
\begin{equation}\label{Exclusive}
S_N = \sum_{i=1}^{N/2} V ({\bf x}_i^{\phantom{x}} - {\bf x}'_{\mathcal{P}(i)}) \ .
\end{equation}
Here, instantons are first grouped into $N/2$ dipoles with $+$ charges at ${\bf x}_i^{\phantom{x}}$ and $-$ charges at ${\bf x}'_i$. Then, the $+$ charge from each dipole $i$ engages in an exclusive interaction with only one $-$ charge from the same or any other dipole $\mathcal{P}(i)$. All possible interaction pairings are captured by the permutations $\mathcal{P}$ and need to be summed over in (\ref{IPartFunct}). While natural for $V(r)\propto r$, this character of interactions is required of any confining potential.

Section \ref{secNonConf} analyzes the non-exclusive interactions (\ref{NonExclusive}) with a functional extension of the Coulomb gas renormalization group \cite{Kosterlitz1977}. Stronger-than-Coulomb and logarithmic potentials generate corrections which become progressively more important at large distances and look like a run-away flow into confinement. The perturbation theory quickly breaks down due to the development of an infra-red divergence. However, an infra-red divergence of the partition function (\ref{IPartFunct}) cannot be cured by any renormalization of the interaction because the underlying coarse-graining and rescaling must preserve the partition function. Instead, the bare potential must be different. Section \ref{secConf} explains how this difference comes about at the more microscopic level where (\ref{IPartFunct}) is derived, and then develops a non-perturbative functional renormalization group for the ensuing exclusive confining interactions (\ref{Exclusive}). Considering also finite temperatures in Section \ref{secFiniteTemp}, we compile a picture of two universality classes for instanton confinement.

\subsection{Non-confining interactions}\label{secNonConf}

Here we consider the attractive interaction potential
\begin{equation}\label{NonConfInt}
C(\delta{\bf x}) \equiv -V(\delta{\bf x}) = -\sum_{n} \frac{K_{n}}{\delta x^n} \quad,\quad K_n>0
\end{equation}
which vanishes at large distances  $\delta x\equiv|\delta{\bf x}|\to \infty$ and has a singularity at $\delta x=0$. This is the Laurent expansion of an arbitrary non-confining potential about $\delta x=0$. The screened potential $C'(\delta x) \equiv C'(|{\bf x}-{\bf x}'|)$ can be determined by inserting a ``test'' instanton dipole at ${\bf x}, {\bf x}'$ and computing
\begin{eqnarray}\label{EffInteraction1}
e^{-C'(\delta x)} &=& \frac{1}{\mathcal{Z}}\sum_{N=0}^{\infty}\frac{y_{0}^{N}}{R_{0}^{DN}}\frac{1}{N_+!N_-!}\int\prod_{i=1}^{N}d^{D}x_{i}\,\exp\biggl\lbrace \\
&& \quad \frac{1}{2}\sum_{i\neq j}q_{i}q_{j}C(|{\bf x}_{i}-{\bf x}_{j}|)-C(|{\bf x}-{\bf x}'|) \nonumber \\
&& \quad +\sum_{i}q_{i}C(|{\bf x}_{i}-{\bf x}|)-\sum_{i}q_{i}C(|{\bf x}_{i}-{\bf x}'|) \nonumber \biggr\rbrace \nonumber
\end{eqnarray}
in the path integral with the instanton action (\ref{NonExclusive}). We use the fugacity as an expansion parameter, assuming $y_0\ll 1$, and restrict to globally neutral instanton configurations, i.e. $N=2k$ and $\sum q_i=0$. Substituting (\ref{IPartFunct}), we find to the lowest order in fugacity:
\begin{eqnarray}\label{EffInteraction2}
&& e^{-C'(\delta x)} e^{C(\delta x)} = 1+\frac{y_{0}^{2}}{R_{0}^{2D}}\int d^{D}R\,d^{D}r\,e^{-C(r)} \\
&& \;\quad\times \left\lbrack e^{C\left(\left|{\bf R}\!+\!\frac{{\bf r}}{2}\!-\!{\bf x}\right|\right)\!-\!C\left(\left|{\bf R}\!-\!\frac{{\bf r}}{2}\!-\!{\bf x}\right|\right)\!-\!C\left(\left|{\bf R}\!+\!\frac{{\bf r}}{2}\!-\!{\bf x}'\right|\right)\!+\!C\left(\left|{\bf R}\!-\!\frac{{\bf r}}{2}\!-\!{\bf x}'\right|\right)}-1\right\rbrack \nonumber
\end{eqnarray}
Here, only one fluctuating dipole with size ${\bf r}={\bf x}_1-{\bf x}_2$ and the center of mass ${\bf R} = \frac{1}{2}({\bf x}_1+{\bf x}_2)$ screens the test dipole. We anticipate that only small fluctuating dipoles make a significant impact, but this must be verified by scrutinizing the $r\to\infty$ limit inside the integral:
\begin{eqnarray}
&& e^{-C'(\delta x)} e^{C(\delta x)} \xrightarrow{r\to\infty}{} 1+\frac{y_{0}^{2}}{R_{0}^{2D}}\int d^{D}R\,d^{D}r \\
&& \;\quad\times \biggl\lbrace -\sum_{n}\frac{2^{2+n}nK_{n}}{r^{1+n}}\,\hat{\bf r}\,\delta{\bf x} + \frac{1}{2}\biggl\lbrack \sum_{n}\frac{2^{2+n}nK_{n}}{r^{1+n}}\,\hat{\bf r}\,\delta{\bf x}\biggr\rbrack^{2} \nonumber \\
&& \qquad +\cdots \biggr\rbrace \nonumber \ .
\end{eqnarray}
Only the quadratic term remains after integrating out the fluctuating dipole orientations $\hat{\bf r}$,
\begin{equation}
e^{-C'(\delta x)} e^{C(\delta x)} \xrightarrow{r\to\infty}{} 1+\frac{y_{0}^{2}}{R_{0}^{2D}}\int dr\,r^d \sum_n \frac{a_n K_n \delta x^2 }{r^{2+2n}} + \cdots \nonumber
\end{equation}
We are not concerned with the $R$ integral because we assumed $|{\bf r}|\gg|{\bf R}|$. But, there is a lower bound
\begin{equation}\label{InfraRed1}
n > \frac{d-1}{2}
\end{equation}
for the acceptable terms in the expansion (\ref{NonConfInt}) which do not cause an infra-red divergence of (\ref{EffInteraction1}). A well-behaved integral is indeed dominated by small $r$, so we estimate its value by Taylor-expanding (\ref{EffInteraction2}) about $r=0$:
\begin{eqnarray}\label{EffInteraction3}
&& e^{-C'(\delta x)} e^{C(\delta x)} = 1+\frac{y_{0}^{2}}{2DR_{0}^{2D}}\int d^{D}R\,d^{D}r\,e^{-C(r)} \\
&& \;\quad\times \left\lbrace r^{2}\Bigl\lbrack\boldsymbol{\nabla}_{{\bf R}}C(|{\bf R}-{\bf x}|)-\boldsymbol{\nabla}_{{\bf R}}C(|{\bf R}-{\bf x}'|)\Bigr\rbrack^{2}+\mathcal{O}(r^{4})\right\rbrace \nonumber \\ \nonumber
\end{eqnarray}
Again, we only keep the $\hat{\bf r}$-independent terms \footnote{Sufficiently high order terms in this expansion create infra-red divergence, but this is an artifact of the expansion. The exact integral is convergent and approximated by its well-behaved lowest order term.}. Next, we turn our attention to the $R$ integral and find
\begin{eqnarray}\label{Rintegral}
&& \int d^{D}R\,\Bigl\lbrack\boldsymbol{\nabla}_{{\bf R}}C(|{\bf R}-{\bf x}|)-\boldsymbol{\nabla}_{{\bf R}}C(|{\bf R}-{\bf x}'|)\Bigr\rbrack^{2} \nonumber \\
&&\qquad = \sum_{n,m} nmK_{n}K_{m}\beta_{n,m}\,\delta x^{D-2-(n+m)} \ .
\end{eqnarray}
Defining the area of a unit $n$-sphere
\begin{equation}
S_n = \frac{2\pi^{(n+1)/2}}{\Gamma\left(\frac{n+1}{2}\right)} \ ,
\end{equation}
the positive dimensionless coefficients $\beta_{n,m}$ are given by
\begin{widetext}
\begin{eqnarray}\label{Beta1}
&& \beta_{n,m}=\frac{S_{D-2}}{2^{D-2-(n+m)}}\int\limits_{0}^{\infty}d\xi\,\xi^{-D+1+(n+m)}\int\limits_{0}^{\pi}d\theta\,\sin^{D-2}\theta \\
&& \quad \times\!\left(\frac{1}{(1+\xi^{2}-2\xi\cos\theta)^{1+(n+m)/2}}\!+\!\frac{1}{(1+\xi^{2}+2\xi\cos\theta)^{1+(n+m)/2}}\!-\!\frac{2(1-\xi^{2})}{(1+\xi^{2}+2\xi\cos\theta)^{1+n/2}(1+\xi^{2}-2\xi\cos\theta)^{1+m/2}}\right)  \nonumber
\end{eqnarray}
\end{widetext}
(note $\xi=\delta x/2R$). The infra-red behavior of $\beta_{n,m}$ obtains by taking the limit $\xi\to 0$ (i.e. $R\to\infty$) inside its integrand:
\begin{eqnarray}
\beta_{n,m}\xrightarrow{\xi\to 0} \textrm{const.} \times \int\limits_{0}^{\infty}d\xi\,\xi^{-D+1+(n+m)} \xi^2 \propto \xi^{4-D+(n+m)} \nonumber
\end{eqnarray}
This imposes another requirement on the acceptable terms in the expansion (\ref{NonConfInt}), less strict than (\ref{InfraRed1}):
\begin{equation}\label{InfraRed2}
n > \frac{D}{2} - 2 = \frac{d-3}{2} \ .
\end{equation}
The integral is well-behaved in the $\xi\to\infty$ limit ($D>0$), but has an ultra-violet singularity for $n+m>D-2$ at $\xi=1$, $\theta\in\lbrace 0,\pi \rbrace$. Substituting $\xi=1+\delta\xi$ and $\theta=\delta\theta$ or $\theta=\pi-\delta\theta$, then expanding to the lowest orders in $\delta\xi, \delta\theta$ and integrating out $\delta\xi \in (\delta\xi_0,\infty)$, $\delta\theta\in(0,\infty)$ reveals:
\begin{eqnarray}\label{Beta2}
&& \beta_{n,m}\xrightarrow[{\sin\theta\to 0}]{\xi\to1} \frac{S_{D-2}\Gamma\left(\frac{D-1}{2}\right)}{2^{D-2-(n+m)}}\,\frac{\Gamma\left(\frac{n+m-D+3}{2}\right)}{\Gamma\left(1+\frac{n+m}{2}\right)} \\
&&\qquad\qquad\qquad\times \frac{\left\vert \delta\xi_{0}\right\vert^{D-2-(n+m)}}{|D-2-(n+m)|}-\textrm{const} \nonumber
\end{eqnarray}
The ultra-violet singularity corresponds to the overlaps between the positions of the virtual and test instantons. Overlapping instantons are physically indistinguishable below the ultra-violet cut-off length scale $R_{0}$, so the integral over $R$ should be regularized by cutting out the volumes of the order of $R_{0}^{D}$ around ${\bf R}\pm\delta{\bf x}/2$. In other words,
\begin{equation}
\xi=\frac{\delta x}{2R} \xrightarrow{R\to\delta x/2} 1+\delta\xi_{0}\sim\frac{\delta x}{\delta x\mp R_{0}}=1\pm\frac{R_{0}}{\delta x} \ .
\end{equation}
So, substituting $\delta\xi_0 \sim R_0/\delta x$ in (\ref{Beta2}) and (\ref{Rintegral}) gives us a qualitatively accurate renormalization (\ref{EffInteraction3}) of the instanton potential to the order of $y_0^2$:
\begin{eqnarray}
&& e^{-C'(\delta x)} e^{C(\delta x)} = 1\!+\!\frac{y_{0}^{2}}{2DR_{0}^{2D}} \sum_{n,m} nmK_{n}K_{m} \!\int\! d^{D}r\,e^{-C(r)} \nonumber \\
&& \qquad\quad\times r^2 \Bigl(\gamma^{\phantom{x}}_{n,m}R_0^{D-2-(n+m)}-\beta'_{n,m}\,\delta x^{D-2-(n+m)}\Bigr) \nonumber \ .
\end{eqnarray}
The coefficients $\beta'_{n,m}$ and $\gamma^{\phantom{x}}_{n,m}$ are positive.

Now we can derive the renormalization group beta functions for the interaction couplings $K_n$. We first coarse-grain the view of dynamics by integrating out fluctuations of small dipoles, $r\in( R_0, e^{\delta l} R_0)$, where $\delta l\ll 1$:
\begin{eqnarray}
&& e^{-C'(\delta x)} e^{C(\delta x)} = 1\!+\!\frac{y_{0}^{2}}{2DR_{0}^{2D}} \sum_{n,m} nmK_{n}^{\phantom{x}} K_{m}^{\phantom{x}} S_{D-1}^{\phantom{x}} R_0^{D+2} \delta l \nonumber \\
&& \qquad\times e^{-C(R_0)} \Bigl(\gamma^{\phantom{x}}_{n,m}R_0^{D-2-(n+m)}-\beta'_{n,m}\,\delta x^{D-2-(n+m)}\Bigr) \nonumber
\end{eqnarray}
Then, to the lowest order in $y_0^2$,
\begin{eqnarray}
&& C'(\delta x) = C(\delta x) - \frac{y_{0}^{2}}{2DR_{0}^{2D}} \sum_{n,m} nmK_{n}^{\phantom{x}} K_{m}^{\phantom{x}} S_{D-1}^{\phantom{x}} R_0^{D+2} \delta l \nonumber \\
&& \qquad\times e^{-C(R_0)} \Bigl(\gamma^{\phantom{x}}_{n,m}R_0^{D-2-(n+m)}-\beta'_{n,m}\,\delta x^{D-2-(n+m)}\Bigr) \nonumber
\end{eqnarray}
It can be formally seen from (\ref{EffInteraction2}) that adding a constant to the potential has no physical effect. Thus, we are free to interpret the entire bracket content in the last expression as the non-confining interaction with power $D-2-(n+m)<0$ (shifted by a constant). Then, substituting the expansion (\ref{NonConfInt}) and matching powers gives us:
\begin{equation}
K'_n = K^{\phantom{x}}_n - \frac{y_{0}^{2}}{R_{0}^{D-2}} \sum_{m} \beta''_{D-2+n-m,m} K_{D-2+n-m}^{\phantom{x}} K_{m}^{\phantom{x}} \delta l \nonumber
\end{equation}
with positive coefficients $\beta''_{n,m}$ for $n,m>0$ limited by (\ref{InfraRed1}). This yields the contribution of coarse-graining to the beta function $(K'_n-K_n)/\delta l \to \delta K_n/\delta l$. After coarse-graining, we rescale the coordinates to restore the original cut-off, $e^{\delta l}R_0 \to R_0$:
\begin{equation}
C'(r)=C(e^{\delta l}r) \quad\Rightarrow\quad \frac{K'_n-K_n}{\delta l} = \frac{\delta K_n}{\delta l} = -nK_n \ .
\end{equation}
Combining the coarse-graining and rescaling gives us the full beta function:
\begin{equation}\label{betaK}
\frac{\delta K_n}{\delta l} = - \frac{y_{0}^{2}}{R_{0}^{D-2}} \sum_{m} \beta''_{D-2+n-m,m} K_{D-2+n-m}^{\phantom{x}} K_{m}^{\phantom{x}} - nK_n \ .
\end{equation}

The renormalization group flow of fugacity $y_0 = e^{-S_0}$ is obtained by fixing the partition function under the rescaling of the ultra-violet cut-off $R_{0}$. The action of a single instanton
\begin{equation}
\mathcal{S}_{1}=\mathcal{S}_{0}+C(\lambda;R_{0}) \ ,
\end{equation}
contains the core part $S_0$ and the ``coherent'' part estimated by the (attractive) instanton interaction potential $C(r)$ at an infra-red cut-off distance $\lambda$. The potential can generally depend on the ultra-violet cut-off $R_0$. The partition function of a single instanton stays invariant under the change of scale $l\to l+\delta l$,
\begin{eqnarray}
&& \mathcal{Z}_{1}\sim\left(\frac{\lambda}{R_{0}}\right)^{D}e^{-\mathcal{S}_{1}} = y_{0}(R_{0})\left(\frac{\lambda}{R_{0}}\right)^{D}e^{-C(\lambda,R_{0})} \nonumber \\
&& \qquad\quad = y_{0}(e^{\delta l}R_{0})\left(\frac{\lambda}{e^{\delta l}R_{0}}\right)^{D}e^{-C(\lambda,e^{\delta l}R_{0})}
\end{eqnarray}
so that
\begin{eqnarray}
&& y_{0}(R_{0}+R_{0}\delta l) = y_{0}(R_{0})+\delta y_{0} \\
&& \qquad\quad = y_{0}(R_{0})\,\left(1+D\delta l+\frac{\partial C(\lambda)}{\partial R_{0}}R_{0}\delta l\right) \ . \nonumber
\end{eqnarray}
This yields the beta function for fugacity
\begin{equation}\label{betaY}
\frac{\delta y_{0}}{\delta l} = y_{0}\left(D+\frac{\partial C(\lambda)}{\partial R_{0}}R_{0}\right) \ .
\end{equation}
The non-confining potential (\ref{NonConfInt}) must be cut off in the ultra-violet limit $r\to 0$. The proper way to do it is to shift it by a constant that achieves $C(R_0)=0$:
\begin{equation}
C(\delta{\bf x}) \to  \sum_{n} K_{n} \left(\frac{1}{R_0^n}-\frac{1}{\delta x^n}\right) \ .
\end{equation}
The implication for the fugacity is
\begin{equation}
\frac{\delta y_{0}}{\delta l} = y_{0}\left(D-\sum_n\frac{nK_n}{R_0^n}\right) \ .
\end{equation}

The full set of renormalization group equations for non-confining instanton interactions is:
\begin{eqnarray}\label{RG1}
\frac{\delta K_n}{\delta l} &=& -\frac{y_{0}^{2}}{R_{0}^{D-2}} \sum_{m} \beta''_{D-2+n-m,m} K_{D-2+n-m}^{\phantom{x}} K_{m}^{\phantom{x}} - nK_n \nonumber \\
\frac{\delta y_{0}}{\delta l} &=& y_{0}\left(D-\sum_n\frac{nK_n}{R_0^n}\right) \ .
\end{eqnarray}
Since $n>0$ and $\beta''_{n,m}>0$, attractive interaction couplings $K_n>0$ generally decrease with the scale parameter $l$, irrespective of the value of fugacity $y_0$. Eventually, sufficiently small $K_n$ cannot prevent the growth of $y_0$. There is only one basin of attraction, $K_n\to 0$, $y_0\to 1(\infty)$. The fixed point corresponds to prolific non-interacting instantons with a vanishing core action $S_0$, and hence represents the deconfined phase.

Even though the last conclusion looks simple, the generation of new interaction channels from the microscopic ones holds important further insight. If interaction channels with $K_n\neq 0$ are limited to $n\in \lbrack M,N \rbrack$, then the first equation in (\ref{RG1}) generates new interactions in channels $n'\in \lbrack 2M+2-D,2N+2-D \rbrack$. Only the generalized Coulomb interaction $n=D-2$ yields a closed set of interactions channels, which can be handled in a more straight-forward manner \cite{Kosterlitz1977}. Weaker-than-Coulomb potentials $n>D-2$ generate an endless set of even weaker non-confining potentials whose main effect is to alter the interactions at short distances. Therefore, the concluded absence of instanton confinement holds consistently for the Coulomb and less confining potentials.

Note that a microscopic attractive interaction in channel $n>D-2$ can formally induce a first-generation repulsive interaction between opposite-charge instantons in a higher channel $n'>n$. Then, the attractive $n$ and repulsive $n'$ channels together generate an attractive second-generation potential in the channel $n''=n+n'-(D-2)>n'$. As the new interaction channels are introduced, the net interaction remains attractive at large distances as dictated by the original lowest channel $n$. The generation of formally repulsive channels is innocuous for two reasons: (i) the net interaction is renormalized only at short distances, and (ii) all the couplings $K_n\to 0$ are irrelevant in the renormalization group sense in all channels.

The most interesting aspect of the interaction renormalization is that stronger-than-Coulomb $n<D-2$ potentials generate an open set of interaction channels $n'<n$ which progressively become more rather than less important at large distances. This breaks down the current approach when the channels violating (\ref{InfraRed1}), or even confining channels $n<0$, are reached. Attempting to still take the equations (\ref{RG1}) seriously reveals a transformation of the $n<0$ interaction couplings $K_n$ into relevant operators which flow away from zero while $y_0\ll 1$. This might indicate an evolution of stronger-than-Coulomb interactions into confining potentials at large distances, but needs to be examined using different methods. Specifically, the infra-red divergence of an interaction potential correction in any renormalization group step likely reveals an infra-red divergence of the partition function (\ref{IPartFunct}) at some order in $y_0$, not necessarily lowest.

\subsection{Logarithmic interactions}\label{secLog}

A special case of interactions between instantons is the logarithmic potential
\begin{equation}
C(r) = -V(r) = K \log\left(\frac{r}{R_0}\right) \ .
\end{equation}
This is a Coulomb potential in $D=2$ space-time dimensions which leads to the Kosterlitz-Thouless transition \cite{Kosterlitz1974} between confined and deconfined phases of instantons. In higher dimensions $D>2$, we must check if an infra-red divergence arises in (\ref{Rintegral}) due to small virtual dipole fluctuations:
\begin{eqnarray}\label{Rlog}
&& \int d^{D}R\,\Bigl\lbrack\boldsymbol{\nabla}_{{\bf R}}C(|{\bf R}-{\bf x}|)-\boldsymbol{\nabla}_{{\bf R}}C(|{\bf R}-{\bf x}'|)\Bigr\rbrack^{2} = \nonumber \\
&& \quad = K^2 \delta x^{2}\int \frac{d^{D}R}{\left(R^{2}+\frac{1}{4}\delta x^{2}\right)^{2}-\left({\bf R}\delta{\bf x}\right)^{2}} \ .
\end{eqnarray}
Power counting shows that divergences from $R\to\infty$ occur in $D\ge 4$. Specifically, in $D=4$ we have the weakest logarithmic divergence:
\begin{eqnarray}\label{Rlog4}
&& \delta x^{2} \int \frac{d^{4}R}{\left(R^{2}+\frac{1}{4}\delta x^{2}\right)^{2}-\left({\bf R}\delta{\bf x}\right)^{2}} = \\
&& \quad = 4\pi^{2}\int\limits _{0}^{\infty}dR\,R\left( 1-\frac{\left|R^2-\frac{1}{4}\delta x^2\right|}{R^{2}+\frac{1}{4}\delta x^{2}}\right) \nonumber \\
&& \quad \xrightarrow{R\to\infty}{} -2\pi^{2}\delta x^{2}\log\left(\frac{\delta x}{R}\right) + \pi^{2}\delta x^{2} + \mathcal{O} \left(\frac{\delta x^2}{R^2}\right) \ . \nonumber
\end{eqnarray}
If the naive renormalization of the microscopic potential appears infra-red divergent, then a different microscopic mechanism must be engaged in the generation and renormalization of interactions. We will discuss this mechanism in the next section. Logarithmic interactions in all $D\ge 4$ space-time dimensions must be treated as confining.

The direct renormalization of a logarithmic interaction in $D=3$ is both infra-red and ultra-violet convergent:
\begin{equation}\label{Rlog3}
\delta x^{2} \int \frac{d^{3}R}{\left(R^{2}+\frac{1}{4}\delta x^{2}\right)^{2}-\left({\bf R}\delta{\bf x}\right)^{2}} = \pi^{3}\delta x \ .
\end{equation}
However, coarse-graining generates a new confining potential $C'(\delta x) \propto \delta x$. This hints a renormalization of the logarithmic interaction into a confining form at macroscopic scales, but also breaks the perturbative renormalization group: further interaction corrections due to the generated linear potential are infra-red divergent. If this corresponds to an infra-red divergence of the partition function, then the resolution is again the confinement we describe in the next section. Otherwise, some earlier studies have suggested that the $D=3$ logarithmic interaction can drive a confinement-deconfinement phase transition \cite{Sudbo2002, Sudbo2004}, while others have argued renormalization into a non-confining Coulomb potential \cite{Herbut2003, Herbut2004}. These works did not transparently consider a possible infra-red divergence.

\subsection{Confining interactions}\label{secConf}

We previously found that instanton interactions which do not decrease fast enough with the distance cause an infra-red divergence of the screening (\ref{EffInteraction1}). This includes algebraic potentials (\ref{NonConfInt}) that violate (\ref{InfraRed1}), and logarithmic potentials in $D\ge 4$ space-time dimensions. The infra-red divergence comes from the wandering of the virtual instanton dipoles' center-of-mass ${\bf R}$ throughout space-time. If the largest possible values of $|{\bf R}|$ produce the most statistically-important instanton configurations for screening, then the virtual dipole fluctuations are effectively suppressed in the thermodynamic limit since they are unlikely to occur anywhere near the test dipole.

A physical system can always find its way out of this trouble. If instanton fluctuations are to take place, their interactions must be restructured by the appropriate dynamics of the system's matter field $\psi$. Instanton interactions are subject to the boundary condition (\ref{InstantonCharge}). One way to satisfy this condition is to evenly spread the flux of the topological current $\mathcal{J}_\mu$ over the spherical manifold $S^d$ which encloses an instanton. If this leads to infra-red divergences, then the flux must spread differently. The alternative extreme option is for the flux to focus into strings. Preserving (\ref{InstantonCharge}) then leads to the exclusive confining two-body interactions ({\ref{Exclusive}) with a linear form $C(r)\propto r$ at large distances. Any non-linear form is either not exclusive, or exhibits some additional bias among the interacting instantons which requires a more complex dynamics. Interestingly, it is possible to calculate the renormalization of exclusive potentials exactly. We will show that the exclusive interaction given by ({\ref{Exclusive}) always resolves the problem of infra-red divergence.

Consider $+$ instantons at locations ${\bf x}_i^{\phantom{x}}$ and an equal number of $-$ instantons at locations ${\bf x}'_i$. We visualize exclusive interactions by connecting every instanton with exactly one other instanton of opposite charge. A connection diagram is specified by a permutation $\mathcal{P}$: the $+$ instanton $i$ is connected to the $-$ instanton $\mathcal{P}(i)$. The instanton partition function with exclusive interactions is:
\begin{eqnarray}\label{ConfPartFunct}
&& \mathcal{Z} = \sum_{N=0}^{\infty}\left(\frac{1}{N!}\right)^{2}\frac{y_{0}^{2N}}{R_{0}^{2ND}} \\
&& \qquad\times \sum_{\mathcal{P}\lbrack N\rbrack}\int\prod_{i=1}^{N}d^{D}x_{i}^{\phantom{x}}\,d^{D}x'_{i}\,\exp\left\lbrack-\sum_{i=1}^{N}C({\bf x}_{i}^{\phantom{x}}-{\bf x}'_{\mathcal{P}(i)})\right\rbrack \ . \nonumber
\end{eqnarray}
When a test dipole is inserted at $({\bf x}_0^{\phantom{x}}, {\bf x}'_0)$, its interaction potential acquires the following renormalization by virtual instanton fluctuations:
\begin{eqnarray}
&& e^{-C'({\bf x}_{0}^{\phantom{x}}-{\bf x}'_{0})} = \frac{1}{\mathcal{Z}}\sum_{N=0}^{\infty}\left(\frac{1}{N!}\right)^{2}\frac{y_{0}^{2N}}{R_{0}^{2ND}} \\
&& \quad\times \sum_{\mathcal{P}\lbrack N+1\rbrack}\int\prod_{i=1}^{N}d^{D}x_{i}^{\phantom{x}}\,d^{D}x'_{i}\,\exp\left\lbrack-\sum_{i=0}^{N}C({\bf x}_{i}^{\phantom{x}}-{\bf x}'_{\mathcal{P}(i)})\right\rbrack \nonumber
\end{eqnarray}
If the instantons of the test dipole are not connected to any of the virtual instantons, we will say that the connection diagram is disconnected. Let us first demonstrate that only the connected diagrams contribute to the interaction renormalization. Consider a generic diagram of $N$ virtual dipoles, $i=1,\dots N$ in which only $k$ dipoles make a connected cluster with the test dipole $i=0$. There are $\binom{N}{k}$ ways to chose $k$ out of $N$ available instantons of each charge to create the connected sub-diagram. Combining with the $1/N!$ factors gives us the sum over connected clusters only ($N=k+m$):
\begin{eqnarray}
&& e^{-C'({\bf x}_{0}^{\phantom{x}}-{\bf x}'_{0})} = 
\frac{1}{\mathcal{Z}}\sum_{k=0}^{\infty}\sum_{m=0}^{\infty}\left(\frac{1}{(k+m)!}\right)^{2}\binom{k+m}{k}^{2} \nonumber \\
&& ~\,\times\! \frac{y_{0}^{2k}}{R_{0}^{2kD}}\!\!\!\sum_{\mathcal{P}\lbrack k+1\rbrack}^{\textrm{conn.}}\int\!\prod_{i=1}^{k}d^{D}x_{i}^{\phantom{x}} d^{D}x'_{i}\,\exp\!\left\lbrack -\!\sum_{i=0}^{k}C({\bf x}_{i}^{\phantom{x}}\!\!-\!{\bf x}'_{\mathcal{P}(i)})\right\rbrack \! \nonumber \\
&& ~\,\times\! \frac{y_{0}^{2m}}{R_{0}^{2mD}}\!\!\sum_{\mathcal{P}\lbrack m\rbrack}\int\!\prod_{j=1}^{m}d^{D}x_{j}^{\phantom{x}} d^{D}x'_{j}\,\exp\!\left\lbrack -\!\sum_{j=1}^{m}C({\bf x}_{j}^{\phantom{x}}\!\!-\!{\bf x}'_{\mathcal{P}(j)})\right\rbrack \! \nonumber \\
&& \, = \sum_{k=0}^{\infty}\left(\frac{1}{k!}\right)^{2}\frac{y_{0}^{2k}}{R_{0}^{2kD}}\sum_{\mathcal{P}\lbrack k+1\rbrack}^{\textrm{conn.}} \nonumber \\
&& \quad\times \int\prod_{i=1}^{k}d^{D}x_{i}^{\phantom{x}} d^{D}x'_{i}\,\exp\left\lbrack -\sum_{i=0}^{k}C({\bf x}_{i}^{\phantom{x}}-{\bf x}'_{\mathcal{P}(i)})\right\rbrack \ .
\end{eqnarray}
There are only two connected clusters with sizes $k=0$ and $k=1$, because the test dipole has only two terminals which can be either connected to each other ($k=0$) or to the terminals of another dipole ($k=1$). Therefore,
\begin{eqnarray}\label{ConfInt}
e^{-C'({\bf x}-{\bf x}')} &=& e^{-C({\bf x}-{\bf x}')} \\
&& +\frac{y_{0}^{2}}{R_{0}^{2D}}\int d^{D}y\,d^{D}y'\,e^{-C({\bf y}-{\bf x}')-C({\bf y}'-{\bf x})} \nonumber
\end{eqnarray}
yields the exact renormalized potential $C'(r)$ to all orders in $y_0$:
\begin{eqnarray}\label{ScreenedPotential}
C'(r) &=& -\log \left( e^{-C(r)}+\alpha^2 y_0^2  \right) \\
&\to&
\begin{cases}
  C(r)-\alpha^2 y_{0}^{2}e^{C(r)} & ,\quad\alpha^2 y_{0}^{2}\ll e^{-C(r)}\\
  -\log(\alpha^2 y_{0}^{2})-\frac{e^{-C(r)}}{\alpha^2 y_{0}^{2}} & ,\quad\alpha^2 y_{0}^{2}\gg e^{-C(r)}
\end{cases} \nonumber
\end{eqnarray}
where
\begin{equation}\label{AlphaDef}
\alpha\lbrack C\rbrack = \frac{1}{R_{0}^{D}} \int d^{D}y\,e^{-C(y)} \ .
\end{equation}

We may consider various forms of the unscreened potential $C(r)$ in order to understand the present approach better. Confining attractive potentials $C(r)\propto r^n$, $n>0$ always keep this procedure infra-red convergent, while the logarithmic attractive potential $C(r)=K\log(r/R_0)$ needs a large enough coupling constant $K>D$ to keep $\alpha$ finite. The renormalized potential $C'$ is not confining even though the microscopic one is. This makes sense only when $C'$ is correctly interpreted as the effective action $C'(r)\equiv S_{\textrm{eff}}(r)$ for an inserted instanton dipole of size $r$. If the dipole starts off small and we gradually increase its size, then for a while we keep paying more and more confinement potential ``energy'' according to $C'(r)\approx C(r)$. The eventual flattening of $C'(r)\to\textrm{const.}$ at sufficiently large $r>\lambda$ indicates that the inserted test instantons become screened. The cross-over from the confining to the flattened regime reveals the screening length $\lambda$:
\begin{equation}\label{ConfLength}
e^{-C(\lambda)}=\alpha^2y_{0}^{2}\quad\Rightarrow\quad C(\lambda)=-\log(\alpha^2y_{0}^{2}) \ .
\end{equation}
We can also define the bare confinement length $\lambda_0$ as the dipole size which provides enough unscreened potential ``energy'' for a new compensating dipole,
\begin{equation}
C(\lambda_0^{\phantom{x}}) = 2S_0^{\phantom{x}} = -\log(y_0^2) \ .
\end{equation}
In comparison to $\lambda_0$, the screening length $\lambda$ is affected by a certain renormalization that enters through the factor of $\alpha^2$ in (\ref{ConfLength}); generally, $\alpha>1$ implies $\lambda<\lambda_0$. While $\lambda_0$ unambiguously increases when the confining interaction $C(r)$ becomes weaker, $\lambda$ can exhibit the opposite behavior for sufficiently weak interactions because $\alpha$ increases at the same time according to (\ref{AlphaDef}). This effect is due to entropy. A strong confining interaction places the compensating instantons very close to the original test instantons. In contrast, a weak confining interaction (more asymptotically free) is much less selective; even though the corresponding $\lambda_0$ is larger, virtual instanton fluctuations have an enhanced probability and contribute significantly to screening, thereby reducing $\lambda$. Note, however, that the present approach cannot be trusted to determine a small $\lambda$ in the weak-interaction limit since the underlying assumption of exclusive interactions need not apply to very short distances.

An important feature of the renormalized potential (\ref{ScreenedPotential}) is that screening occurs inevitably at a finite length scale $\lambda$ regardless of whether the unscreened potential $C(r)$ is truly confining in the $r\to\infty$ limit. Exclusive confining interactions between instantons are formed beyond some short length-scale, and may extend to a finite distance $\xi$, given for example by the coherence length of the fundamental field $\psi$, as long as $\xi>\lambda$. If circumstances reduce $\xi$ below $\lambda$, instanton confinement will be lost. Similarly, thermal deconfinement requires a finite temperature scale in the presence of asymptotic freedom.

Our main goal is to identify the conditions for confinement phase transitions and explore their universal properties. The usual tool for this purpose is renormalization group, but here we have to slightly depart from the tradition and construct a stripped-down scaling argument. The scaling $\lambda\to 0$ of the screening length will serve as an indicator of confinement, and computing $\lambda$ with (\ref{ConfLength}) requires the unscreened potential $C(r)$, not the renormalized one. At the same time, the fugacity $y_0$ is determined by the full partition function, so it implicitly depends on the renormalized potential. The general beta function for fugacity (\ref{betaY}) involves the scale-dependent renormalized potential only through its singular dependence on the ultra-violet cut-off length $R_0$. The key observation for our purposes is that all screening renormalizations alter only the large-distance behavior of the interaction potential $C(r)$, so the renormalized potential has the same short-distance singularities as the unscreened one. Therefore, we may use the unscreened potential $C(r)$ in the beta function (\ref{betaY}) for fugacity.

We will track the unscreened potential and fugacity as functions of the observation scale parametrized by $l\ge 0$. Coordinate rescaling modifies the potential according to
\begin{equation}
C'({\bf r})=C({\bf r})+\delta C({\bf r})=C(e^{\delta l}{\bf r})=C({\bf r})+{\bf r}\boldsymbol{\nabla}C({\bf r})\,\delta l \nonumber
\end{equation}
which amounts to $\delta C({\bf r})/\delta l={\bf r}\boldsymbol{\nabla}C({\bf r})$. No further renormalization of the unscreened potential takes place under coarse-graining. For completeness, the full set of scaling equations for isotropic potentials is:
\begin{equation}\label{RG2}
\frac{\delta C(r)}{\delta l} = r \frac{\partial C(r)}{\partial r} 
\quad,\quad
\frac{\delta y_{0}}{\delta l} = y_{0}\left(D+\frac{\partial C(\lambda)}{\partial R_{0}}R_{0}\right) \ .
\end{equation}
These equations are accurate to all orders of $y_0$, so we can reliably track the flow of fugacity to large values. They also apply to arbitrary confining functions $C(r;l;R_0)$ which do not create infra-red divergences in the present approach. The exact solution for the potential scaling is
\begin{equation}
C(r;l) = C(re^l) \ .
\end{equation}

It turns out that identifying fixed points is not particularly useful because they lie outside of the limits in which the scaling equations are valid. Formally, $C(r)=\textrm{const.}$ is a fixed point potential, but it breaks down this procedure by introducing an infra-red divergence in (\ref{AlphaDef}). Fugacity always has $y_0=0$ and $y_0\to1(\infty)$ fixed points. The logarithmic potential $C(r)=K\log(r/R_0)$ scales in a manner $C(r;l)=K\log(e^l r/R_0)$ which keeps the interaction strength $K$ fixed and allows a non trivial fixed point for fugacity at $K=D$, but this and smaller couplings also make (\ref{AlphaDef}) divergent.

Instead of considering fixed points and their stability, we can analyze the evolution of concrete potentials. It is easy to see that a generic asymptotically-free power-law potential becomes only more confining at larger length scales
\begin{equation}
C(r) = \sum_{n>0} A_n r^n \quad\Rightarrow\quad C(r;l)=\sum_{n>0}A_{n}e^{nl}r^{n} \ ,
\end{equation}
without jeopardizing asymptotic freedom. Since this interaction has no ultra-violet singularities, the fugacity scales as
\begin{equation}
\frac{dy_{0}}{dl} = D y_{0}\quad\Rightarrow\quad y_{0}(l)=y_{0}(0)\,e^{Dl}
\end{equation}
This is unbounded flow toward infinity, but we should stop it at $y_0=1$ because fugacity $y_0 = e^{-S_0}$ is given by the action $S_0$ of an instanton core, so $y_0\to 1$ corresponds to the vanishing cost of a core. If we focus on a single channel $n$, i.e. $C(r)=A_n r^n$, we can also easily compute (\ref{AlphaDef})
\begin{equation}
\alpha(l) = \frac{S_{d}\,\Gamma\!\left(\frac{D}{n}\right)}{nR_{0}^{D}A_{n}^{D/n}}\,e^{-Dl}
\end{equation}
and solve (\ref{ConfLength}) to find the running screening length
\begin{equation}\label{RunningScreening}
\lambda(l) = \lambda(0)\,e^{-l}
\end{equation}
in terms of the microscopic screening length
\begin{equation}
\lambda(0) = 
\begin{cases}
\left\lbrack \frac{2}{A_{n}}\left\vert \log\bigl\lbrack y_{0}(0)\,\alpha(0)\rbrack\right\vert \right\rbrack^{1/n} & ,\quad y_{0}(0)\,\alpha(0)<1\\
\;R_{0}\to0 & ,\quad y_{0}(0)\,\alpha(0)>1
\end{cases}
\end{equation}
Recall that the formal saturation $\lambda\to R_0$ in the $\alpha y_0>1$ regime cannot be trusted because it pertains to weak interactions which need not assume an exclusive form at short distances. The main insight for now is that instantons are necessarily confined at $T=0$ according to $\lambda(l)\to0$ as $l\to\infty$.

We can repeat this exercise for logarithmic potentials
\begin{equation}
C(r;l) = K\log\left(\frac{r}{R_{0}}e^{l}\right) \ .
\end{equation}
The new ingredient is an ultra-violet singularity from $R_0\to0$ that enables a non-trivial flow of fugacity:
\begin{equation}
\frac{dy_{0}}{dl} = (D-K)y_{0}\quad\Rightarrow\quad y_{0}(l)=y_{0}(0)\,e^{(D-K)l} \ .
\end{equation}
Only the cases $K>D$ are accessible without infra-red divergences in the present approach. The fugacity then flows into the $y_0\to0$ fixed point and pushes the screening length $\lambda$ toward infinity,
\begin{equation}
\lambda(l)=e^{-l}R_{0}\left(e^{(2K-D)l}\frac{K-D}{S_{d}\,y_{0}(0)}\right)^{2/K} \xrightarrow{l\to\infty} \infty \ .
\end{equation}
Instantons are formally deconfined, but their fluctuations are suppressed by $y_0\to 0$ and the photons of the gauge field (\ref{PhotonCorrel}) have an infinite correlation length at $T=0$. At finite temperatures, the absence of confinement makes the correlation length finite. We will find indications that the state with $K<D$ is deconfined and short-range correlated even at zero temperature in the next section.

An exclusive logarithmic potential considered here is not realistic. However, the last analysis reveals a general possibility of a non-trivial $y_0\to 0$ fixed point shaped by the short-distance modifications of a physical confining potential. Note that a crossover to a non-exclusive interaction at short distances does not by itself jeopardize this argument because (\ref{betaY}) holds for non-exclusive potentials as well. It is not obvious what microscopic dynamics may pull the fugacity toward zero. Nevertheless, instantons' positional fluctuations can alter the effect of (\ref{InstantonCharge}) at short distances from the instanton, and accommodate the potential. Overall, the behavior of fugacity determines two different universality classes of instanton suppression at $T=0$, characterized by the fixed points $y_0\to0$ and $y_0\to1(\infty)$.

\subsection{Instanton confinement at finite temperatures}\label{secFiniteTemp}

Here we analyze the fate of instanton confinement at finite temperatures $T$ by developing a two-stage renormalization group. In stage 1, we start from a $D=d+1$ dimensional space-time which hosts the imaginary-time path integral (\ref{PartFunct}) of a $d$-dimensional quantum system at a finite temperature $T$. The time-direction has periodic boundary conditions and a finite extent $\tau\in(0,\beta)$, where $\beta=T^{-1}$ is the inverse temperature. The system remains infinite in all $d$ spatial directions, and the Lagrangian density of the system is the same as in the $T=0$ case. The renormalization group proceeds exactly as in the previous section, but the span of the imaginary time coordinate changes under scaling as:
\begin{equation}\label{BetaScaling}
\frac{dT}{dl}=lT\quad\Rightarrow\quad T(l)=T(0)\,e^{l}\quad,\quad\beta(l)=\beta(0)\,e^{-l} \ .
\end{equation}
We must stop the stage 1 at a finite scale $l=l_0$ when $\beta\to R_{0}$ becomes comparable to the ultra-violet cut-off length and the system effectively flattens to its $d$-dimensional spatial manifold. The running coupling constants have picked by then the essential quantum renormalizations, while the pure scaling of the energy couplings $w(l)=w(0) e^l$ has imparted on them a factor of $\beta$. Then, we embark on the ``classical'' stage 2 of renormalization group, utilizing the last $d$-dimensional action of $\beta=R_{0}$ which has a Landau-Ginzburg form. The final fixed point at $l\to\infty$ signifies the phase in which the microscopic theory lives. The entire procedure is depicted in Fig.\ref{FigTRG}.

\begin{figure}
\subfigure[{}]{\includegraphics[width=1.6in]{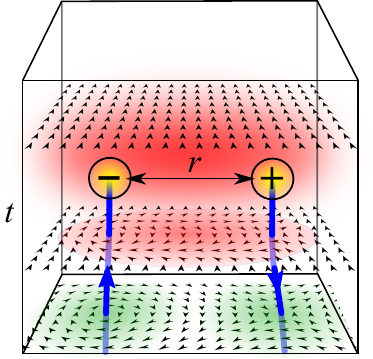}}\hspace{0.1in}
\subfigure[{}]{\includegraphics[width=1.6in]{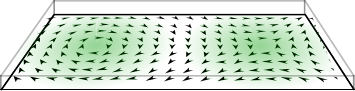}}
\caption{\label{FigTRG}(a) Stage 1 of the renormalization group: the temporal (vertical) extent of the world $\beta=1/T$ flows toward zero. While $\beta$ is large enough, the configurations of topological defects and instantons can comfortably fit into the world view. In particular, the field configuration (e.g. the shown phase gradient $\boldsymbol{\nabla}\theta$) can maintain a confining action potential $V(r)\propto r$ between the instantons. Such confined instantons are macroscopically irrelevant if their confinement length $\lambda$ scales down to zero during this stage. (b) Stage 2 of the renormalization group: $\beta$ has scaled down to the cut-off length, and the world has lost its imaginary time dimension. If the instantons have not been confined under $\lambda\to 0$ yet, they cannot be properly defined any more. Any surviving spatial gradients, which build topological defects, can take their minimum-action form. This typically produces an algebraic $U(r)\propto r^{-\alpha}$ potential energy between the defects ($\alpha>0$). One could interpret $U(r)$ as the interaction potential between instanton remnants, but this is not enough for confinement. An exception is $U(r)\propto\log(r)$ in $d=2$ spatial dimensions, which permits a Kosterlitz-Thouless transition.}
\end{figure}

The following discussion assumes that the coherence length $\xi$ of the microscopic field $\psi$ remains larger than the instanton screening length $\lambda$ at all scales $l$. During stage 1, the confining interaction potential $C(r)$ and fugacity $y_{0}$ flow according to (\ref{RG2}). Let us first consider the asymptotically free regime $y_{0}\to1(\infty)$. The confinement length $\lambda$ and the inverse temperature $\beta$ both decrease with the scale parameter and race toward the cut-off $R_0$. If $\lambda$ reaches the cut-off first, the instantons become strictly confined and neutralized at microscopic scales during the stage 1. Further scale evolution of dynamics features only the fluctuations of neutral dipole excitations, since the running theory cannot keep any information about the microscopic details below the cut-off length $R_0$. Dipoles are all that's left when the stage 2 begins and we merely expect that their positive mass drives the remaining renormalization group flow as a relevant operator. The final fixed point is confined.

An alternative to this outcome is the scenario in which $\beta$ reaches $R_0$ sooner than $\lambda$. If that happens, then individual instanton charges, or more accurately their projections, are still discernible in the $d$-dimensional world view at the end of stage 1. However, these instanton projections cannot interact with confining potentials in the resulting $d=D-1$ dimensional world. Confining interactions were derived in the first place from the action cost of the stiff matter field $\psi$. Let us assume that $\psi$ is coherent across some length scale $\xi$. Consider a pair of opposite-charge topological defects coexisting at the same instant of time $\tau$ a spatial distance $\delta r < \xi$ apart. They can be simultaneously removed by a pair of instantons at time $\tau_0$. In order to minimize action, a coherent $\psi$ field configuration corresponding to this removal must involve a deformation of $\psi({\bf r}, \tau)$ from the (locally) optimal spatial configuration at $\tau<\tau_0-\delta\tau$ to a costly spatial configuration at time $\tau=\tau_0$ which is uniform everywhere except inside a narrow flux tube that connects the defect and antidefect. The instanton pair abruptly removes the tube and replaces it with the surrounding uniform $\psi$ at times $\tau>\tau_0$. This was illustrated in Fig.\ref{FigInst}(a). The action cost proportional to $\delta r$ comes from the rapid temporal variations $(\partial_0 \psi)^2$ along the tube, but the required gradient terms are expelled from the purely spatial $d$-dimensional action of the stage 2. This obviously cannot be repaired by non-simultaneous instantons, so no mechanism is left to maintain confining interactions between instantons. Ultimately, instantons are not well-defined in stage 2. Their projections assume at best a dynamics analogous to that of spatial topological defects; the strength of interactions is limited by the Coulomb law $C(r)\sim r^{2-d}$ (at least in the case of monopoles and hedgehogs \cite{Nikolic2019}), which is not able to confine its charges (see Section \ref{secNonConf}). The final fixed point is deconfined.

Since two different fixed points of renormalization group can be reached at macroscopic $l\to\infty$ scales, a thermodynamics phase transition marks the resulting sharp distinction between the confined and deconfined phases of instantons. Asymptotically free confining interactions $C(r)=A_n r^n$ in the $y_0\to 1(\infty)$ universality class produce a deconfinement phase transition at a critical temperature $T_{c}>0$. Given (\ref{RunningScreening}), (\ref{BetaScaling}) and the above discussion, the critical temperature is
\begin{equation}
T_{c}=\frac{1}{\lambda(0)} = \left\lbrack \frac{2}{A_{n}}\left\vert\, \log\left(y_{0}(0)\,\frac{S_{d}\,\Gamma\!\left(\frac{D}{n}\right)}{nR_{0}^{D}A_{n}^{D/n}}\right)\right\vert \right\rbrack^{-1/n} \ .
\end{equation}
The critical temperature is generally finite, but goes to zero if $y_{0}\to0$ at any fixed $A_{n}$. It increases with the interaction strength $A_n$ when $A_n$ is large enough. Formally, $T_c$ has a local minimum at a small value of $A_n=\widetilde{A}_n$ and also grows with decreasing $A_n<\widetilde{A}_n$, but again, this weak-interaction regime pushes the theory out of its applicability limits. We can also define a critical screening length at temperature $T$
\begin{equation}
\lambda_{c}(T)=\frac{1}{T} \ ,
\end{equation}
such that $\lambda(0)<\lambda_{c}(T)$ yields a confined phase.

In the absence of a Berry curvature, the photon correlation length $\zeta$ introduced in Section \ref{secTDyn} exhibits a critical behavior near the confinement phase transition of the $y_{0}\to1(\infty)$ universality class. The exact screened potential (\ref{ScreenedPotential}) and the stage 1 renormalization group show that instanton dipoles are statistically relevant in $T=0$ confined phases only if their sizes are limited by the screening length $\lambda$, i.e.  $r<\lambda$. Conversely, the dipoles with sizes $r>\lambda$ are statistically irrelevant. At finite temperatures, we find from the renormalization group that the dipoles of microscopic sizes $r<\lambda_{c}=1/T$ remain confined in the macroscopic point of view, while the microscopically large $r>\lambda$ dipoles continue to be statistically irrelevant because they are still neutralized in spatial directions by the confining potential of the microscopic theory, i.e. they are confined in the microscopic point of view. Therefore, the statistically important dipoles which contribute to the volume-law behavior of the correlation function (\ref{Wilson2}) are only those whose microscopic sizes are $\lambda_{c}<r<\lambda$. Consider an instanton placed at the origin, and let $P(r)$ be the microscopic cumulative probability that this instanton is neutralized by an anti-instanton within the radius $r$ as a result of the dynamics at $T=0$. We know that $P(0)=0$ and $P(\infty)=1$. Continuity and the absence of other length scales in the problem also ensure that $P(r)$ has no extrema on $0<r<\infty$. At finite temperatures, an anti-instanton placed at $\lambda>r>\lambda_{c}$ will end up being deconfined from the instanton. The probability of this occurrence when $\Delta T=T-T_{c}\ll T_{c}$ is:
\begin{equation}
P(\lambda)-P(\lambda_{c}) \approx P'(\lambda_{c})\,(\lambda-\lambda_{c})\approx \lambda_{c}P'(\lambda_{c})\,\frac{\Delta T}{T_{c}} \ .
\end{equation}
This is a conditional probability, assuming that an instanton was present at the origin in the first place (a compensating anti-instanton is then required). The probability density of satisfying the condition can be estimated as $R_{0}^{-D}y_{0}^{\phantom{x}}$. Therefore, the probability per unit space-time volume of finding a free deconfined instanton is roughly:
\begin{equation}
p \approx R_{0}^{-D}y_{0}^{\phantom{x}}\lambda_{c}^{\phantom{x}}P'(\lambda_{c}^{\phantom{x}})\,\frac{\Delta T}{T_{c}} \ .
\end{equation}
We may interpret this as the inverse volume $p\sim\zeta^{-D}$ which contains only one free instanton. Then, the mean distance between uncompensated instantons is
\begin{equation}
\zeta \sim m^{-1} \sim R_{0}^{\phantom{x}}\left\lbrack y_{0}^{\phantom{x}}\lambda_{c}^{\phantom{x}}P'_{\lambda}(\lambda_{c}^{\phantom{x}})\,\frac{\Delta T}{T_{c}}\right\rbrack ^{-1/D} \ .
\end{equation}
The photons of the effective gauge field can propagate coherently only across distances where they do not encounter uncompensated instantons. Therefore, the last equation estimates the photons' correlation length $\zeta$ and their effective mass $m$. We see that the correlation length diverges as a power-law on the approach to the critical temperature. It should be noted, however, that a phase transition with different critical properties can take place if the coherence length $\xi$ of the fundamental field limits the screening length $\lambda$ due to a temperature increase. Analyzing such a scenario requires a more comprehensive theory, so we will not attempt it here.

The alternative $y_0\to 0$ universality class is qualitatively different. We have seen that this regime sends the confining length $\lambda$ toward infinity at $T=0$. Consequently, $\lambda$ never reaches the cut-off $R_0$ and $\beta$ trivially wins the race in the stage 1 of renormalization group. Only a deconfined phase exists any finite temperature. Since the fugacity reaches a finite value $y_0>0$ at the end of stage 1, thermal fluctuations are always able to generate some instantons. The scale parameter $l$ at which the stage 1 ends determines the length scale beyond which instantons effectively lose their confinement. By relating this to temperature, we can find the crossover temperature $T^*$ below which the suppression of instantons approximately conserves the topological charge. Let us first compute the photon correlation length $\zeta$. If the fugacity scales toward zero as $y_0(l)=y_0(0)e^{\kappa l}$, then
\begin{equation}
y_{0}(l)=y_{0}(0)\,\left(R_{0}T\right)^{\kappa} \ .
\end{equation}
at the end of stage 1 where $\beta(l)=R_{0}$. The space-time volume that typically contains a single instanton is proportional to the inverse probability $y_0=e^{-S_{0}}$ of a bare instanton occurrence. Since photons propagate freely only until they encounter an instanton, this volume limits the correlation length $\zeta$ of photons,
\begin{equation}
\zeta^{D}\sim \frac{R_{0}^{D}}{y_{0}(l)} \quad\Rightarrow\quad\zeta\sim\frac{R_{0}}{y_{0}^{1/D}\,\left(R_{0}T\right)^{\kappa/D}} \ .
\end{equation}
The scale $\zeta$ computed this way pertains to the beginning of the second renormalization group stage; all quantum effects have been taken into account, and any further renormalizations would be calculated as in the standard classical renormalization group in $d$ spatial dimensions. The photon mass affected by the quantum processes is $m\sim\zeta^{-1}$.

In the case of a logarithmic potential $C(r)=K\log(r/R_0)$, we have $\kappa = K-D$. The correlation length diverges for $K>D$ when the temperature approaches zero. Supposing that this expression could be trusted for $K<D$ as well, its prediction $\zeta\to 0$ on the approach to zero temperature is consistent with the deconfined instanton plasma. The low-temperature regime with an approximate topological charge conservation is bounded by a crossover temperature $T^*$. Setting $\zeta(T^{*}) = x R_0$, we find
\begin{equation}
T^{*}\sim\frac{1}{R_{0}}\Bigl(x\,y_{0}^{1/D} \Bigr)^{-D/\kappa} \propto R_0^{-1}\,y_0^{-1/(K-D)} \ .
\end{equation}
At $T<T^*$, photon coherence is still discernible at least at some finite short length scales. We must pick a value for $x$ such that $xy_{0}^{1/D}>1$ in order to also get $T^{*}=0$ when $K=D$. Ultimately, $T^{*}$ can be made large by a large instanton interaction coupling $K>D$, together with a naturally small ultra-violet cut-off length $R_{0}$. In this sense, the low-temperature dynamics may be significantly influenced by the quantum phase transition between the states of conserved and unconserved topological charge.

\section{Experimental signatures: noise correlations}\label{secExp}

The phase transition between instanton confined and deconfined phases need not involve any latent heat or symmetry breaking. The question is, then, if it is possible to experimentally detect this transition. Here we argue that noise correlations, measurable in the specific heat, provide a probe for instanton confinement.

The practical feasibility of detailed experimental detection rests upon the means to indirectly observe the fluctuations of topological charge in the system. For example, since vortices are attached to particles in quantum Hall liquids, one can gain insight about vortex dynamics by measuring electron density fluctuations. A relativistic example is given by the chiral quantum anomaly of quantum electrodynamics: the relevant instantons are the creations and annihilations of inter-linked flux loops that carry the Hopf invariant of the $\pi_3(S^2)$ homotopy group, but it is their link to the electron chiral currents that provides the means for detection. The fluctuations driven by instanton dynamics should be visible in the specific heat as well, although with much less detail.

Let us assume for simplicity that we have some indirect way of measuring the total topological charge $N(t)$ inside the entire system at any time $t$. In order to theoretically characterize the fluctuations of $N(t)$, we want to consider the generalized Wilson correlation function (\ref{Wilson2}) on the boundary $S^d$ of a slab which stretches infinitely in all spatial dimensions but has a finite temporal extent within the interval $t \in (t_1,t_2)$. The noise spectrum $\textrm{Var}\,\mathcal{C}(\omega)$ of this correlation function can distinguish between the confined and deconfined instanton phases. Let
\begin{equation}\label{Times}
\bar{t}=\frac{t_{2}+t_{1}}{2}\quad,\quad\delta t=t_{2}-t_{1} \ .
\end{equation}
Then,
\begin{widetext}
\begin{eqnarray}
\delta\mathcal{C}(\delta t) &=& N(t_{2})-N(t_{1}) = \int d\omega\,(e^{i\omega t_{2}}-e^{i\omega t_{1}})N(\omega) = 2i\int d\omega\,e^{i\omega\bar{t}}\sin\left(\frac{\omega\delta t}{2}\right)N(\omega)
\end{eqnarray}
relates the correlation function to the noise spectrum $N(\omega)$ of the topological charge in the system. The variance of these random fluctuations is:
\begin{eqnarray}
&& \textrm{Var}\,\mathcal{C}(\delta t) = \langle\delta\mathcal{C}^2(\delta t)\rangle = \left\langle \left\vert \int d\omega\,(e^{i\omega t_{2}}-e^{i\omega t_{1}})N(\omega)\right\vert ^{2}\right\rangle -\left\vert \left\langle \int d\omega\,(e^{i\omega t_{2}}-e^{i\omega t_{1}})N(\omega)\right\rangle \right\vert ^{2} \\
&& \quad = 2\int d\omega_{1}d\omega_{2}\,e^{i(\omega_{1}-\omega_{2})\bar{t}}\left\lbrack \cos\left(\frac{(\omega_{2}-\omega_{1})\delta t}{2}\right)-\cos\left(\frac{(\omega_{2}+\omega_{1})\delta t}{2}\right)\right\rbrack \Bigl\lbrack\left\langle N(\omega_{1})N^*(\omega_{2})\right\rangle -\left\langle N(\omega_{1})\right\rangle \left\langle N^*(\omega_{2})\right\rangle \Bigr\rbrack \ . \nonumber
\end{eqnarray}
\end{widetext}
The noise spectrum should not depend on the overall times $\bar{t}$ at which the experiment is conducted, so we may take the average over $\bar{t}$ without loss of information:
\begin{equation}
\frac{1}{\Delta\bar{t}} \int d\bar{t}\,\textrm{Var}\,\mathcal{C}(\delta t) = \frac{8\pi}{\Delta\bar{t}}\int d\omega\,\sin^{2}\left(\frac{\omega\delta t}{2}\right)\,\textrm{Var}\Bigl\lbrace|N(\omega)|\Bigr\rbrace \nonumber
\end{equation}
The remaining dependence on $\delta t$ is different in the confined ``area-law'' and deconfined ``volume-law'' regimes. The slab volume enclosed by the $S^d$ manifold, on which we calculate $\mathcal{C}$, is linearly proportional to $\delta t$. In contrast, the area of $S^d$ has a negligible dependence on $\delta t$ in an infinite system. These behaviors restrict the frequency dependence of the variance of $N(\omega)$:
\begin{equation}\label{NoiseSpec}
\textrm{Var}\Bigl\lbrace|N(\omega)|\Bigr\rbrace = 
\begin{cases}
  A\,\delta(\omega)+f(\omega) & ,\quad\textrm{confined}\\
  B\,|\omega|^{-2}+g(\omega) & ,\quad\textrm{deconfined}
\end{cases}
\end{equation}
where
\begin{equation}
f(\omega), g(\omega)=
\begin{cases}
  0 & |\omega|<\delta t_{\textrm{min}}^{-1}\\
  \neq0 & \textrm{otherwise}
\end{cases} \ .
\end{equation}
The arbitrary functions $f,g$ specify the high-frequency spectrum and affect the noise correlations only on short time intervals $\delta t < \delta t_{\textrm{min}}$. They have no bearing on the thermodynamic properties observed in the $S^d\to\infty$, i.e. $\delta t\to\infty$ limit. Note that the ``constants'' $A$ and $B$ can depend on temperature and other parameters. Additional sources of noise in the indirect measurements of $N(t)$ can blur the distinction between these frequency behaviors, but the depletion of low-frequency noise remains a characteristic signature of the confined instanton phase.

If neither direct nor indirect measurements of $N(t)$ can be performed, it is still possible to see the change of noise correlations in the heat capacity. Generally, heat capacity at temperature $T$ is related to the internal energy $\mathcal{E}$ fluctuations:
\begin{equation}\label{HeatCap}
C(T)=\frac{\textrm{Var}\lbrace\mathcal{E}\rbrace}{k_{B}T^{2}} \ .
\end{equation}
Creating a topological defect in an instanton event, perhaps together with an attached particle excitation, always costs some energy $\epsilon$. Therefore, the relevant energy fluctuations should be proportionally derived from the fluctuations of the topological charge $N(t)$. The total internal energy $\mathcal{E}$ inherits such topological fluctuations from a range of low frequencies in the spectrum:
\begin{eqnarray}
\textrm{Var}\lbrace\mathcal{E}\rbrace &\sim& \frac{\epsilon^2}{\Delta\bar{t}}\!\int\limits_{\omega_0}^{\delta t_{\textrm{min}}^{-1}}\! d\omega\,\textrm{Var}\lbrace|N(\omega)|\rbrace \\
&\sim& \frac{\epsilon^2}{\Delta\bar{t}}
\begin{cases}
  A(T) & ,\quad\textrm{confined}\\
  \frac{B(T)}{\omega_{0}} & ,\quad\textrm{deconfined}
\end{cases} \nonumber
\end{eqnarray}
A lower frequency bound $\omega_0$ is formally necessary for the deconfined phase in order to avoid infra-red divergence (we naively set $\omega_0=0$ for the confined phase). The high-temperature behavior $C(T)\propto (\epsilon/T)^2$ expected for uncorrelated gap-$\epsilon$ degrees of freedom means that $\omega_0$ cannot be limited by temperature. Also, the fluctuations of $N$ are governed by an action instead of a Hamiltonian, so there is no clear energy scale either that could determine $\omega_0$. In fact, the only available physical source for $\omega_0$ is the (long) time interval during which the noise statistics is measured. This experiment duration limits both $\bar{t}$ and $\delta t$ in (\ref{Times}), so we expect $\omega_0 \Delta\bar{t} \sim 1$. Since the noise statistics is accurately collected only in the $\Delta\bar{t}\to\infty$ limit, we conclude that instanton deconfinement produces a heat capacity jump at the critical temperature $T_c$:
\begin{equation}
C_{\textrm{inst.}}(T) \propto
\begin{cases}
  0 & ,\quad T<T_c\;\;\textrm{(confined)}\\
  (\epsilon/T)^2 & ,\quad T>T_c\;\;\textrm{(deconfined)}
\end{cases}
\end{equation}

\section{Discussion}\label{secDiscussion}

A prominent non-relativistic realization of instanton confinement is topological order. A prerequisite for topological order is the conservation of topological charge, i.e. instanton confinement. It was observed recently that monopoles and hedgehogs in higher spatial dimensions $d>2$ can seemingly exist in topologically ordered phases at finite temperatures \cite{Nikolic2019}, because their degenerate ground states on topologically non-trivial manifolds are separated by free energy barriers that grow as $L^{d-2}$ with the system size $L$. This is similar to the barriers between the symmetry-related degenerate ground states which enable spontaneous symmetry breaking in macroscopic systems. The present analysis supports finite-temperature topological order by demonstrating instanton confinement at finite temperatures.

Fractional quantum Hall states live in $d=2$ spatial dimensions and their topological order, i.e. ground state degeneracy, is protected only at zero temperature by the finite free energy barriers. However, this only relates to ``vacuum'' instantons and their ability to mix classical topological sectors on topologically non-trivial manifolds. Vacuum instantons create or annihilate topological defects whose singularities \emph{do not} live on the spatial manifold (e.g. vortices threaded through the 2D-world's torus openings). Hence, they are softer than the ``excitation'' instantons on trivial manifolds, which we explore here. Excitation instantons create vortex cores on the manifold, and can maintain confinement at finite temperatures. From this perspective, a thermodynamic sharpness of fractional quantum Hall liquids may extend to low finite temperatures, probably with some aspects of fractionalization surviving as long as both charge and vorticity are conserved.

Pseudogap states have a long history in the context of high-temperature superconductors, and also can be considered in cold atom gases. A pseudogap state may be smoothly connected to a disordered high-temperature phase. A simple example is the crossover from a band-insulator of weakly interacting fermions to a Mott insulator of tightly bound $s$-wave Cooper pairs \cite{Nikolic2010b}, driven by the interaction strength. Such a crossover parallels the BCS-BEC crossover in cold atom gases; there is no phase transition between the two types of insulators, but the Mott regime is characterized by bosonic (instead of fermionic) lowest energy excitations and a bosonic mean-field or XY (instead of BCS) universality class for the superfluid transition \cite{moon:230403, Nikolic2010}. More intricate possibility is a correlated pseudogap phase, thermodynamically distinct from conventional disordered states. Any state of confined instantons is a pseudogap candidate because it is sharply characterized at least by its instanton confinement via (\ref{Wilson2}). Many experiments have revealed distinctive short-range correlations and charge coherence in the underdoped pseudogap state of cuprates \cite{Corson1999, Yazdani2004, Fang2004, Steiner2005, Valla2006, Armitage2007, Hudson2007, Kohsaka2007, Gomes2007, Kohsaka2008, Ghir2012, Torchinsky2013}, above the superconducting critical temperature $T_c$ and below the doping-dependent ``pseudogap'' temperature $T^*$. Among these, Nernst effect measurements \cite{Ong2001, Wang2006, Li2010} find a plausible interpretation in the picture of protected and mobile vortices which drift due to a temperature gradient and produce a voltage drop in the direction perpendicular to their drift. One may ask how vortices can survive when the superconducting order parameter loses long-range coherence. The present study provides a rigorous explanation, and (re)opens the possibility that a phase transition associated with instanton deconfinement happens at the $T^*$ temperature.

A finite density of topological charge is not required for instanton confinement. Therefore, correlated phases with conserved topological charge are also possible in relativistic systems. Such phases can be relativistic analogues of topological orders, although it is not clear yet how to precisely characterize them.

Several interesting systems of particles with strong spin-orbit coupling in $d=2$ and $d=3$ spatial dimensions exhibit periodic arrays of topological defects and anti-defects in their mean-field ground states \cite{Nikolic2011a, Nikolic2014, Nikolic2014a, Nikolic2019b}. The spin-orbit coupling provides a ``magnetic'' length that determines the separation between the defects. One imagines that quantum fluctuations could melt such a defect lattice. If that happens, some defects and anti-defects could annihilate, but a finite density of each would still be energetically protected, even if the defects became delocalized. It is not obvious if such a quantum liquid would possess topological order because the average numbers of defects and anti-defects could be the same. However, the present study shows that this quantum liquid can have a zero \emph{conserved} topological charge, and thus be distinct from a Mott insulator.

Still, there is no guarantee that a non-condensed confined instanton phase is a particular unconventional phase in a particular physical system. To illustrate this point, we can think about bosons living on a lattice and interacting with a U(1) gauge field. Interactions can stabilize a Mott insulator, whereby the boson field $\psi$ becomes incoherent at microscopic scales. A dual picture of a Mott insulator is a vortex or monopole superconductor, so the appropriate topological charge is not conserved. States with instanton confinement and the topological charge conservation which accompanies it, therefore, are not Mott insulators. Any low temperature phase in which the bosons are not confined or localized is a state of confined instantons. This statement can be extended to fermions at zero temperature as well, i.e. the conventional states such as Fermi liquids, integer quantum Hall states and ordinary band insulators. So, the question is if there is anything interesting about instanton confinement in weakly correlated systems.

For example, it is possible that instanton confinement can be related to quantum anomalies in relativistic systems. The chiral quantum anomaly of quantum electrodynamics in $D=4$ space-time dimensions links the non-conservation of the chiral current to the Hopf $\pi_3(S^2)$ topological invariant of the electromagnetic field. This invariant, actually, characterizes instantons because the spatial field configurations can only realize singularity-free topological objects with a Hopf index in $d=3$ spatial dimensions, analogous to skyrmions in $d=2$ dimensions. Even though local quantum tunneling can readily change the Hopf index of these configurations, instanton confinement corresponds to the dynamics which effectively preserves the topological index of the ground state. So, perhaps, the chiral anomaly is a correlation effect which replaces topological order in the non-relativistic systems of massless Dirac particles. After all, it does require interactions via a gauge field. But then, the main question is if sufficient correlations can be achieved at low temperatures in some Dirac systems to sharply distinguish a non-trivial quantum-anomalous phase from the completely uncorrelated high-temperature phase.

Of course, the interesting confined-instanton quantum liquids may be routinely precluded by conventional phase transitions out of the coherent (e.g. Higgs) phase of the matter field. At least a dedicated length scale must be provided in the Hamiltonian for the finite correlation length $\xi$ if the confined-instanton phase is to be possible. If, instead, the model defines only the lattice constant, or some equivalent ultra-violet cut-off length, then the matter field cannot be short-range coherent across any relevant finite distances $\xi>\lambda$. This, in fact, keeps the phase diagram of the compact XY model simple and without ``pseudogap'' phases. If the XY model is gauged and the external magnetic field specifies a magnetic length, then quantum Hall liquids become possible.

Instanton confinement is easiest to understand in the context of bosons because condensates spontaneously break a symmetry and manifestly confine the instantons. But, how does instanton confinement occur in the case of fermions? Due to the Pauli exclusion, fermions cannot maintain a macroscopic phase coherence in their currents. Nevertheless, their quantum coherence is still evident at zero temperature in the simple phenomena like the sharp Fermi surface in interacting Fermi liquids and zero conductivity in band insulators. These features characterize the thermodynamic limit, but degrade at any finite temperature: the Fermi-Dirac distribution of occupation numbers becomes a continuous function of energy, and the conductivity of band-insulators becomes thermally activated. A more complicated phenomenon is the electronic quantum Hall liquid at zero temperature, which can be fractionalized. This type of a state still requires instanton confinement for the quantization of the filling factor $\nu$. How is this possible? We find an answer by observing that delocalized fermions at zero temperature still possess quantum coherence up to the mean inter-particle separation distance $l$, limited by the Pauli exclusion. If $\xi\sim l$ is sufficiently larger than the lattice constant, and the topological defects are separated by comparable distances, we have the basic condition for instanton confinement just as in the case of short-range coherent bosons.

\section{Conclusions}\label{secConclusions}

The main finding of this analysis is that correlated quantum phases of confined instantons can exist at low finite temperatures without spontaneous symmetry breaking. Instanton confinement is a synonym for the conservation of topological charge, i.e. the invariant which characterizes topological defects. Long range coherence of the matter field is not a prerequisite for these phases. The presented renormalization group mathematically reveals that coherence needs to extend only up to the instanton screening distance $\lambda$. Therefore, confined phases can be stabilized with bosonic and fermionic particles alike. In the case of fermions at zero temperature, $\lambda$ must be smaller than the mean distance between particles.

Instanton deconfinement transitions can occur at finite temperatures and exhibit critical properties of second order transitions even though spontaneous symmetry breaking does not take place. This universality class characterizes asymptotically free instantons. We computed the critical scaling of the emergent gauge field's correlation length above the critical temperature, assuming that particles remain coherent across sufficiently long distances. At the same time, we found that an alternative universality class characterizes instantons which are not asymptotically free: topological charge can be conserved only at zero temperature, but the influence of this non-trivial quantum dynamics can be seen below a finite crossover temperature.

These findings generalize quantum liquids of topological defects to finite temperatures and beyond topological order. They may be relevant for the physics of some unconventional phases in correlated materials, such as quantum Hall and spin liquids, pseudogap state of cuprates, and others.

\section{Acknowledgements}\label{secAck}

This research was partly supported by the Department of Energy, Basic Energy Sciences, Materials Sciences and Engineering Award DE-SC0019331. Support was also provided by the Quantum Science and Engineering Center at George Mason University. \\

{\bf Note added:} During the review of this paper, a preprint on a closely related topic was posted by Radzihovsky and Toner (\verb|arXiv:2401.04761 v1|). Their study finds that all stronger-than-Coulomb interactions between charged particles in a plasma ultimately renormalize into Coulomb interactions at large distances. This result contradicts our findings in the case of \emph{confining} interactions, and the disagreement can be attributed to the ignored infra-red divergence of the partition function in their analysis. At the same time, their analysis provides an alternative approach to strong \emph{non-confining} interactions, where the perturbative renormalization group presented here is inconclusive.

\appendix

\section{Instantons for the $\pi_n(S^n)$ homotopy groups}\label{app1}

Spinor fields $\psi$ which transform in a representation of the Spin($d$) group in $d$ spatial dimensions generally admit singular configurations that can be characterized by $\pi_n(S^n)$ homotopy groups \cite{Nikolic2019}. Separate singularities exist in charge and spin sectors. At the lowest rank, the $\pi_1(S^1)$ singularities are vortices in the charge sector or skyrmions in the spin sector; these are point topological defects in $d=2$, lines or loops in $d=3$, two-dimensional sheets in $d=4$, etc. At the highest rank, the $\pi_{d-1}(S^{d-1})$ singularities are always topologically protected point-like monopoles in the charge sector and hedgehogs in the spin sector. At any rank $n<d$, the dynamics of charge-sector singularities can be captured by the Lagrangian density with gradient ($\kappa_n$) and Maxwell ($e_n$) terms
\begin{eqnarray}\label{RankLagrangian}
\mathcal{L}_n &=& \frac{\kappa_n}{2} \left( \sum_{i=1}^{n} (-1)^{i-1} \partial_{\mu_i} A_{\mu_1\cdots\mu_{i-1}\mu_{i+1}\cdots\mu_n} + A_{\mu_1\cdots\mu_n} \right)^2 \nonumber \\
&& + \frac{1}{2e_n^2} (\epsilon_{\mu_1\cdots\mu_{d-n}\nu\lambda_1\cdots\lambda_n} \partial_\nu A_{\lambda_1\cdots\lambda_n})^2 \ ,
\end{eqnarray}
where $A_{\mu_1\cdots\mu_n}$ is an antisymmetric rank-$n$ tensor gauge field whose flux evaluated on closed $n$-manifolds is the quantized topological charge of the enclosed $\pi_n(S^n)$ singularities. The gauge field at rank $n-1$ serves as a matter field minimally coupled to the rank-$n$ gauge field. Analogous hierarchy exists in the spin sector, but the gauge fields at lower ranks have non-Abelian form.

Only monopoles and hedgehogs at the highest rank $d-1$ can enjoy topological protection. Consider the Higgs phase of the matter field at rank $d-1$. The suppressed fluctuations of the matter field quantize the rank $d-1$ flux
\begin{equation}
\frac{1}{q} \oint\limits_{S^{d-1}} d^{d-1}x\, \epsilon_{\mu_1\cdots\mu_{d-1}} A_{\mu_1\cdots\mu_{d-1}} \in \mathbb{Z}
\end{equation}
which emanates from the monopole or hedgehog singularity, where $q$ is the flux quantum ($q=2\pi$ in the charge sector, and $q=S_{n}$ is the area of a unit-radius $n$-sphere in the spin sector). The gauge field, then, depends on the distance $r$ from the singularity as $A_{\mu_1\cdots\mu_{d-1}} \sim 1/r^{d-1}$. The Maxwell term of (\ref{RankLagrangian}) vanishes everywhere except at the singularity, so its cost is local. The gradient term can be compensated via the Higgs mechanism if the matter field embeds the singularity in its configuration, otherwise the energy cost
\begin{equation}
\frac{\kappa_{d-1}}{2} \!\! \int\limits_{|{\bf x}|<R}\!\! d^dx \left(\frac{1}{r^{d-1}}\right)^2 \propto \frac{1}{R^{d-2}}
\end{equation}
with an infra-red cut-off $R$ would impart the Coulomb interaction between static monopoles separated by the distance $R$, with potential energy $U(R)\sim 1/R^{d-2}$. Note that a static monopole is a quantized worldline stretching in the temporal direction of the $D=d+1$ dimensional space-time. Now consider instantons which create or annihilate a point topological defect at time $t=0$. Their action potential $V(r)$ is largest when the matter field at rank $d-1$ is in the Higgs phase. Since the matter field must compensate a singularity, an instanton needs to change its $\pi_{d-1}(S^{d-1})$ topological invariant. This cannot be done with a smooth transformation. The matter field will be discontinuous across $t=0$ and the question is only what time-dependent configuration minimizes the discontinuity action to form an instanton. As explained in the introduction (Section \ref{secTDyn}), the best option is to focus the flux into singular worldlines. The annihilation of a monopole is just a deflection of its world line from the temporal into a spatial direction, which generalizes the Faraday's law of electrodynamics. The gauge flux is ultimately conserved and the monopoles/hedgehogs are topologically protected. The interaction between instantons is linear in the distance between them, $V(r)\propto r$, because it obtains from the string tension (Maxwell term) associated with the flux worldline that connects them. Note that we can trust this mean-field type of conclusion without much concern in $d>2$ dimensions because we did assume a Higgs phase.

None of the finite-sized singular structures at lower ranks are topologically protected. A singular manifold at any rank $n<d-1$ must be without a boundary if it is to have a finite size. An example is a vortex loop in $d=3$ spatial dimensions, which becomes a vortex 2-sphere in $d=4$, etc. This disturbs the gauge fields at all ranks only locally. One can, actually, carry out a smooth deformation of the fields to collect all energy density into a finite volume around the singularity. Specifically, the rank-$n$ gauge field $A_{\mu_1\cdots\mu_n}$ has zero flux on generic $n$-manifolds which ``encircle'' the singularity, because they are not interlinked with the manifold of the singularity (only the rank $d-1$ is special since there the ``encircling'' manifold for the flux calculation is by definition interlinked with the point-manifold of the singularity). Therefore, the most dangerous rank-$n$ gauge field can be made to vanish in all directions a finite distance away from the singularity. Since all energy cost can be consumed into a finite volume, an instanton which removes the finite boundary-less singularity also costs a finite action. The action interaction potential between instantons can only be short-ranged, so the instantons cannot be confined and the number of lower-rank singularities cannot be conserved. It should be emphasized that this concerns quantum fluctuations; instantons are quantum tunneling processes. It doesn't help to interlink a singular structure with something else -- this creates an energy barrier for the removal of the singularity, but instanton quantum tunneling still occurs with a finite probability. The only possible protection against tunneling at ranks $n<d-1$ is to make the singular domains infinite.

The instanton deconfinement at lower ranks $n<d-1$ is even more pronounced in the spin sector where the gauge fields at these ranks are non-Abelian. See the next Appendix for a brief example from the Yang-Mills gauge theory.

\section{Action interaction potential for skyrmion and hopfion instantons}\label{app2}

Certain topological defects do not have a singularity. This makes them more vulnerable to quantum tunneling. The best known example are skyrmions. In $d=2$ spatial dimensions, skyrmions are topological defects of the unit-magnitude vector spin configuration $\hat{\bf n}=(\hat{n}^x,\hat{n}^y,\hat{n}^z)$ in the continuum limit. If one defines a gauge field from the spin chirality
\begin{equation}\label{SkyrmionFlux}
\epsilon_{\mu\nu\lambda}\partial_{\nu}A_{\lambda}=\frac{1}{2}\epsilon_{\mu\nu\lambda}\epsilon^{abc}\hat{n}^{a}(\partial_{\nu}\hat{n}^{b})(\partial_{\lambda}\hat{n}^{c}) \ ,
\end{equation}
then the topological invariant of the spin configuration is the total flux
\begin{equation}\label{SkyrmionCharge}
\frac{1}{4\pi} \oint\limits_{S^2} d^2x\, \hat{\bf z} (\boldsymbol{\nabla}\times{\bf A}) \in \mathbb{Z}
\end{equation}
of the gauge field over the entire space. This number counts the total skyrmion charge (number) in the system. The spatial manifold must, actually, be a sphere $S^2$ in order to obtain quantization, but this is in practice arranged on the infinite open plane by insisting that $\hat{\bf n}$ be uniform and ferromagnetic in far-away regions. In physical terms, $A_\mu$ is the gauge field which imparts ``topological'' Hall effect on electrons that move in the topologically non-trivial magnetic background. A skyrmion does not have a singularity, but it does have a center where all gradients of the magnetization field can be focused by smooth transformations. Once this is done, a tunneling event which removes the skyrmion costs only a finite action. Finite-action instantons interact via short-range action potentials, so they cannot be confined. Skyrmion charge conservation is violated by quantum tunneling events, and hence skyrmions cannot produce a truly quantized ``topological'' Hall effect.

Most generally, an instanton with creates or annihilates a skyrmion is a hedgehog of $\hat{\bf n}$ in space-time. This can be seen by taking the difference between the skyrmion charge $N(t)$ computed from (\ref{SkyrmionCharge}) at two different times $t_1\neq t_2$. The space-time boundary at times $t_1$ and $t_2$ can be continuously deformed into a 2-sphere $S^2$ embedded in space-time, and then (\ref{SkyrmionCharge}) becomes the hedgehog topological index from the $\pi_2(S^2)$ homotopy group. In the worst-case scenario, the interaction between two hedgehogs a space-time distance $r$ apart is given by the Coulomb potential $V(r)\sim1/r$ in $D=d+1=3$. This is also unable to stabilize a confined phase of instantons \cite{Kosterlitz1977}.

One might naively suspect that topological defects without a singularity cannot be protected against quantum fluctuations. However, this is not the case. The vector field $\hat{\bf n}$ supports another type of topological defects in $d=3$ spatial dimensions, usually called hopfions. The corresponding topological invariant can again be expressed using the gauge field from (\ref{SkyrmionFlux}),
\begin{equation}\label{HopfIndex}
\frac{1}{16\pi^2} \int\limits_{S^3} d^3x \, {\bf A}(\boldsymbol{\nabla}\times{\bf A}) \in \mathbb{Z} \ .
\end{equation}
This is the Hopf index of the $\pi_3(S^2)$ homotopy group. Geometrically, hopfions are interlinked loops of skyrmions, which themselves are lines in $d=3$, so they do not contain a singularity. Now, set up a 3-sphere $S^3$ in space-time and evaluate (\ref{HopfIndex}) on it. If $S^3$ is smoothly deformed into a slab between times $t_1$ and $t_2$, then the computed integer is the number of hopfions that were created (or destroyed) in the time interval $(t_1, t_2)$. Thus, (\ref{HopfIndex}) gives us the instanton charge. By dimensional analysis, we see that a local non-zero instanton charge generates a gauge field $A_\mu\sim 1/r$ a (large) distance away from the source. This disturbance cannot be compressed into a finite volume because we need both a non-zero gauge field and its magnetic flux at any distance $r$. So, in the best case scenario, the Maxwell energy cost of a single Hopf instanton scales as
\begin{equation}
\mathcal{L}\propto(\epsilon_{ijk}\partial_{j}A_{k})^{2}+(\partial_{0}A_{i})^{2}\sim\frac{1}{r^{4}}
\end{equation}
\begin{equation}
S(R)=\mathcal{S}_{0}+\int\limits_{R_{0}}^{R}dr\!\!\int\limits_{S^{3}(r)}\!\!d^{3}x\,\mathcal{L} \sim S_{0}+\log\left(\frac{R}{R_{0}}\right)
\end{equation}
with the system radius $R$. Therefore, Hopf instantons interact at least via a logarithmic potential $\log(r/R_0)$. We found in Section \ref{secLog} that this interaction cannot apply to every pair of instantons in $D=4$ space-time dimensions. Instead, it evolves into an exclusive interaction at large distances, linear in the distance between the instantons. Consequently, Hopf index is globally conserved at sufficiently low temperatures despite the quantum tunneling fluctuations.

This conclusion holds for the Abelian gauge field $A_\mu$. However, the Hopf index can be also introduced for non-Abelian gauge fields $A_\mu^{\phantom{x}} = A_\mu^a \gamma^a$, where $\gamma^a$ are the gauge group generators. Defining
\begin{equation}
F_{\mu\nu} = \partial_\mu A_\nu - \partial_\nu A_\mu - i \lbrack A_\mu, A_\nu \rbrack \quad,\quad \widetilde{F}_{\mu\nu} = \frac{1}{2} \epsilon_{\mu\nu\alpha\beta} F_{\alpha\beta} \nonumber
\end{equation}
(the square brackets are a commutator), we evaluate the Hopf index on the closed ``surface'' $S^3$ by integrating over the space-time volume $B^4$ bounded by it \cite{ZinnJustin2001}:
\begin{equation}
\frac{1}{32\pi^2} \int\limits_{B^4} d^4x\; F_{\mu\nu}^a \widetilde{F}_{\mu\nu}^a \in \mathbb{Z} \ .
\end{equation}
Even though smooth transformations of the gauge field protect this integer, there are singular instanton configurations which alter it with only a finite action cost -- at least in the $D=4$ Yang-Mills gauge theory \cite{Belavin1975, tHooft1976, Shuryak1998, tHooft2000, Oglivie2012}. Their interaction potential is short-ranged in the best case scenario, and the ensuing instanton deconfinement is responsible for quark confinement in QCD.



\begin{thebibliography}{88}%
\makeatletter
\providecommand \@ifxundefined [1]{%
 \@ifx{#1\undefined}
}%
\providecommand \@ifnum [1]{%
 \ifnum #1\expandafter \@firstoftwo
 \else \expandafter \@secondoftwo
 \fi
}%
\providecommand \@ifx [1]{%
 \ifx #1\expandafter \@firstoftwo
 \else \expandafter \@secondoftwo
 \fi
}%
\providecommand \natexlab [1]{#1}%
\providecommand \enquote  [1]{``#1''}%
\providecommand \bibnamefont  [1]{#1}%
\providecommand \bibfnamefont [1]{#1}%
\providecommand \citenamefont [1]{#1}%
\providecommand \href@noop [0]{\@secondoftwo}%
\providecommand \href [0]{\begingroup \@sanitize@url \@href}%
\providecommand \@href[1]{\@@startlink{#1}\@@href}%
\providecommand \@@href[1]{\endgroup#1\@@endlink}%
\providecommand \@sanitize@url [0]{\catcode `\\12\catcode `\$12\catcode
  `\&12\catcode `\#12\catcode `\^12\catcode `\_12\catcode `\%12\relax}%
\providecommand \@@startlink[1]{}%
\providecommand \@@endlink[0]{}%
\providecommand \url  [0]{\begingroup\@sanitize@url \@url }%
\providecommand \@url [1]{\endgroup\@href {#1}{\urlprefix }}%
\providecommand \urlprefix  [0]{URL }%
\providecommand \Eprint [0]{\href }%
\providecommand \doibase [0]{http://dx.doi.org/}%
\providecommand \selectlanguage [0]{\@gobble}%
\providecommand \bibinfo  [0]{\@secondoftwo}%
\providecommand \bibfield  [0]{\@secondoftwo}%
\providecommand \translation [1]{[#1]}%
\providecommand \BibitemOpen [0]{}%
\providecommand \bibitemStop [0]{}%
\providecommand \bibitemNoStop [0]{.\EOS\space}%
\providecommand \EOS [0]{\spacefactor3000\relax}%
\providecommand \BibitemShut  [1]{\csname bibitem#1\endcsname}%
\let\auto@bib@innerbib\@empty
\bibitem [{\citenamefont {Wen}(2004)}]{WenQFT2004}%
  \BibitemOpen
  \bibfield  {author} {\bibinfo {author} {\bibfnamefont {X.-G.}\ \bibnamefont
  {Wen}},\ }\href@noop {} {\emph {\bibinfo {title} {{Quantum Field Theory of
  Many-Body Systems}}}}\ (\bibinfo  {publisher} {Oxford University Press},\
  \bibinfo {address} {New York},\ \bibinfo {year} {2004})\BibitemShut {NoStop}%
\bibitem [{\citenamefont {Wen}(2017)}]{Wen2017}%
  \BibitemOpen
  \bibfield  {author} {\bibinfo {author} {\bibfnamefont {X.-G.}\ \bibnamefont
  {Wen}},\ }\href {\doibase 10.1103/RevModPhys.89.041004} {\bibfield  {journal}
  {\bibinfo  {journal} {Reviews of Modern Physics}\ }\textbf {\bibinfo {volume}
  {89}},\ \bibinfo {pages} {041004} (\bibinfo {year} {2017})}\BibitemShut
  {NoStop}%
\bibitem [{\citenamefont {Levin}\ and\ \citenamefont
  {Stern}(2009)}]{Levin2009}%
  \BibitemOpen
  \bibfield  {author} {\bibinfo {author} {\bibfnamefont {M.}~\bibnamefont
  {Levin}}\ and\ \bibinfo {author} {\bibfnamefont {A.}~\bibnamefont {Stern}},\
  }\href@noop {} {\bibfield  {journal} {\bibinfo  {journal} {Physical Review
  Letters}\ }\textbf {\bibinfo {volume} {103}},\ \bibinfo {pages} {196803}
  (\bibinfo {year} {2009})}\BibitemShut {NoStop}%
\bibitem [{\citenamefont {Maciejko}\ \emph {et~al.}(2010)\citenamefont
  {Maciejko}, \citenamefont {Qi}, \citenamefont {Karch},\ and\ \citenamefont
  {Zhang}}]{Maciejko2010}%
  \BibitemOpen
  \bibfield  {author} {\bibinfo {author} {\bibfnamefont {J.}~\bibnamefont
  {Maciejko}}, \bibinfo {author} {\bibfnamefont {X.-L.}\ \bibnamefont {Qi}},
  \bibinfo {author} {\bibfnamefont {A.}~\bibnamefont {Karch}}, \ and\ \bibinfo
  {author} {\bibfnamefont {S.-C.}\ \bibnamefont {Zhang}},\ }\href@noop {}
  {\bibfield  {journal} {\bibinfo  {journal} {Physical Review Letters}\
  }\textbf {\bibinfo {volume} {105}},\ \bibinfo {pages} {246809} (\bibinfo
  {year} {2010})}\BibitemShut {NoStop}%
\bibitem [{\citenamefont {Swingle}\ \emph {et~al.}(2011)\citenamefont
  {Swingle}, \citenamefont {Barkeshli}, \citenamefont {McGreevy},\ and\
  \citenamefont {Senthil}}]{Swingle2011}%
  \BibitemOpen
  \bibfield  {author} {\bibinfo {author} {\bibfnamefont {B.}~\bibnamefont
  {Swingle}}, \bibinfo {author} {\bibfnamefont {M.}~\bibnamefont {Barkeshli}},
  \bibinfo {author} {\bibfnamefont {J.}~\bibnamefont {McGreevy}}, \ and\
  \bibinfo {author} {\bibfnamefont {T.}~\bibnamefont {Senthil}},\ }\href@noop
  {} {\bibfield  {journal} {\bibinfo  {journal} {Physical Review B}\ }\textbf
  {\bibinfo {volume} {83}},\ \bibinfo {pages} {195139} (\bibinfo {year}
  {2011})}\BibitemShut {NoStop}%
\bibitem [{\citenamefont {Walker}\ and\ \citenamefont
  {Wang}(2012)}]{Walker2012}%
  \BibitemOpen
  \bibfield  {author} {\bibinfo {author} {\bibfnamefont {K.}~\bibnamefont
  {Walker}}\ and\ \bibinfo {author} {\bibfnamefont {Z.}~\bibnamefont {Wang}},\
  }\href@noop {} {\bibfield  {journal} {\bibinfo  {journal} {Frontiers of
  Physics}\ }\textbf {\bibinfo {volume} {7}},\ \bibinfo {pages} {150} (\bibinfo
  {year} {2012})}\BibitemShut {NoStop}%
\bibitem [{\citenamefont {von Keyserlingk}\ \emph {et~al.}(2013)\citenamefont
  {von Keyserlingk}, \citenamefont {Burnell},\ and\ \citenamefont
  {Simon}}]{Keyserlingk2013}%
  \BibitemOpen
  \bibfield  {author} {\bibinfo {author} {\bibfnamefont {C.~W.}\ \bibnamefont
  {von Keyserlingk}}, \bibinfo {author} {\bibfnamefont {F.~J.}\ \bibnamefont
  {Burnell}}, \ and\ \bibinfo {author} {\bibfnamefont {S.~H.}\ \bibnamefont
  {Simon}},\ }\href {\doibase 10.1103/PhysRevB.87.045107} {\bibfield  {journal}
  {\bibinfo  {journal} {Physical Review B}\ }\textbf {\bibinfo {volume} {87}},\
  \bibinfo {pages} {045107} (\bibinfo {year} {2013})}\BibitemShut {NoStop}%
\bibitem [{\citenamefont {Jian}\ and\ \citenamefont {Qi}(2014)}]{Jian2014}%
  \BibitemOpen
  \bibfield  {author} {\bibinfo {author} {\bibfnamefont {C.-M.}\ \bibnamefont
  {Jian}}\ and\ \bibinfo {author} {\bibfnamefont {X.-L.}\ \bibnamefont {Qi}},\
  }\href {\doibase 10.1103/PhysRevX.4.041043} {\bibfield  {journal} {\bibinfo
  {journal} {Physical Review X}\ }\textbf {\bibinfo {volume} {4}},\ \bibinfo
  {pages} {041043} (\bibinfo {year} {2014})}\BibitemShut {NoStop}%
\bibitem [{\citenamefont {Wang}\ and\ \citenamefont {Levin}(2014)}]{Wang2014a}%
  \BibitemOpen
  \bibfield  {author} {\bibinfo {author} {\bibfnamefont {C.}~\bibnamefont
  {Wang}}\ and\ \bibinfo {author} {\bibfnamefont {M.}~\bibnamefont {Levin}},\
  }\href@noop {} {\bibfield  {journal} {\bibinfo  {journal} {Physical Review
  Letters}\ }\textbf {\bibinfo {volume} {113}},\ \bibinfo {pages} {080403}
  (\bibinfo {year} {2014})}\BibitemShut {NoStop}%
\bibitem [{\citenamefont {Ye}\ and\ \citenamefont {Gu}(2015)}]{Ye2015}%
  \BibitemOpen
  \bibfield  {author} {\bibinfo {author} {\bibfnamefont {P.}~\bibnamefont
  {Ye}}\ and\ \bibinfo {author} {\bibfnamefont {Z.-C.}\ \bibnamefont {Gu}},\
  }\href {\doibase 10.1103/PhysRevX.5.021029} {\bibfield  {journal} {\bibinfo
  {journal} {Physical Review X}\ }\textbf {\bibinfo {volume} {5}},\ \bibinfo
  {pages} {021029} (\bibinfo {year} {2015})}\BibitemShut {NoStop}%
\bibitem [{\citenamefont {Ye}\ and\ \citenamefont {Gu}(2016)}]{Ye2016}%
  \BibitemOpen
  \bibfield  {author} {\bibinfo {author} {\bibfnamefont {P.}~\bibnamefont
  {Ye}}\ and\ \bibinfo {author} {\bibfnamefont {Z.-C.}\ \bibnamefont {Gu}},\
  }\href@noop {} {\bibfield  {journal} {\bibinfo  {journal} {Physical Review
  B}\ }\textbf {\bibinfo {volume} {93}},\ \bibinfo {pages} {205157} (\bibinfo
  {year} {2016})}\BibitemShut {NoStop}%
\bibitem [{\citenamefont {Putrov}\ \emph {et~al.}(2017)\citenamefont {Putrov},
  \citenamefont {Wang},\ and\ \citenamefont {Yau}}]{Putrov2017}%
  \BibitemOpen
  \bibfield  {author} {\bibinfo {author} {\bibfnamefont {P.}~\bibnamefont
  {Putrov}}, \bibinfo {author} {\bibfnamefont {J.}~\bibnamefont {Wang}}, \ and\
  \bibinfo {author} {\bibfnamefont {S.-T.}\ \bibnamefont {Yau}},\ }\href
  {\doibase 10.1016/j.aop.2017.06.019} {\bibfield  {journal} {\bibinfo
  {journal} {Annals of Physics}\ }\textbf {\bibinfo {volume} {384}},\ \bibinfo
  {pages} {254} (\bibinfo {year} {2017})}\BibitemShut {NoStop}%
\bibitem [{\citenamefont {Fuji}\ and\ \citenamefont
  {Furusaki}(2019)}]{Furusaki2019}%
  \BibitemOpen
  \bibfield  {author} {\bibinfo {author} {\bibfnamefont {Y.}~\bibnamefont
  {Fuji}}\ and\ \bibinfo {author} {\bibfnamefont {A.}~\bibnamefont
  {Furusaki}},\ }\href {\doibase 10.1103/PhysRevB.99.241107} {\bibfield
  {journal} {\bibinfo  {journal} {Physical Review B}\ }\textbf {\bibinfo
  {volume} {99}},\ \bibinfo {pages} {241107} (\bibinfo {year}
  {2019})}\BibitemShut {NoStop}%
\bibitem [{\citenamefont {Lan}\ \emph {et~al.}(2018)\citenamefont {Lan},
  \citenamefont {Kong},\ and\ \citenamefont {Wen}}]{Wen2019a}%
  \BibitemOpen
  \bibfield  {author} {\bibinfo {author} {\bibfnamefont {T.}~\bibnamefont
  {Lan}}, \bibinfo {author} {\bibfnamefont {L.}~\bibnamefont {Kong}}, \ and\
  \bibinfo {author} {\bibfnamefont {X.-G.}\ \bibnamefont {Wen}},\ }\href@noop
  {} {\bibfield  {journal} {\bibinfo  {journal} {Physical Review X}\ }\textbf
  {\bibinfo {volume} {8}},\ \bibinfo {pages} {021074} (\bibinfo {year}
  {2018})}\BibitemShut {NoStop}%
\bibitem [{\citenamefont {Lan}\ and\ \citenamefont {Wen}(2019)}]{Wen2019}%
  \BibitemOpen
  \bibfield  {author} {\bibinfo {author} {\bibfnamefont {T.}~\bibnamefont
  {Lan}}\ and\ \bibinfo {author} {\bibfnamefont {X.-G.}\ \bibnamefont {Wen}},\
  }\href@noop {} {\bibfield  {journal} {\bibinfo  {journal} {Physical Review
  X}\ }\textbf {\bibinfo {volume} {9}},\ \bibinfo {pages} {021005} (\bibinfo
  {year} {2019})}\BibitemShut {NoStop}%
\bibitem [{\citenamefont {Berezinsky}(1972)}]{Berezinsky1972}%
  \BibitemOpen
  \bibfield  {author} {\bibinfo {author} {\bibfnamefont {V.~L.}\ \bibnamefont
  {Berezinsky}},\ }\href@noop {} {\bibfield  {journal} {\bibinfo  {journal}
  {Soviet Physics JETP}\ }\textbf {\bibinfo {volume} {34}},\ \bibinfo {pages}
  {610} (\bibinfo {year} {1972})},\ \bibinfo {note} {zh. Eksp. Teor, Fiz. 61,
  1144 (1972).}\BibitemShut {Stop}%
\bibitem [{\citenamefont {Kosterlitz}\ and\ \citenamefont
  {Thouless}(1973)}]{Kosterlitz1973}%
  \BibitemOpen
  \bibfield  {author} {\bibinfo {author} {\bibfnamefont {J.~M.}\ \bibnamefont
  {Kosterlitz}}\ and\ \bibinfo {author} {\bibfnamefont {D.~J.}\ \bibnamefont
  {Thouless}},\ }\href@noop {} {\bibfield  {journal} {\bibinfo  {journal}
  {Journal of Physics C: Solid State Physics}\ }\textbf {\bibinfo {volume}
  {6}},\ \bibinfo {pages} {1181} (\bibinfo {year} {1973})}\BibitemShut
  {NoStop}%
\bibitem [{\citenamefont {Kosterlitz}(1974)}]{Kosterlitz1974}%
  \BibitemOpen
  \bibfield  {author} {\bibinfo {author} {\bibfnamefont {J.~M.}\ \bibnamefont
  {Kosterlitz}},\ }\href@noop {} {\bibfield  {journal} {\bibinfo  {journal}
  {Journal of Physics C: Solid State Physics}\ }\textbf {\bibinfo {volume}
  {7}},\ \bibinfo {pages} {1046} (\bibinfo {year} {1974})}\BibitemShut
  {NoStop}%
\bibitem [{\citenamefont {Wilson}(1974)}]{Wilson1974}%
  \BibitemOpen
  \bibfield  {author} {\bibinfo {author} {\bibfnamefont {K.~G.}\ \bibnamefont
  {Wilson}},\ }\href {\doibase 10.1103/PhysRevD.10.2445} {\bibfield  {journal}
  {\bibinfo  {journal} {Physical Review D}\ }\textbf {\bibinfo {volume} {10}},\
  \bibinfo {pages} {2445} (\bibinfo {year} {1974})}\BibitemShut {NoStop}%
\bibitem [{\citenamefont {Polyakov}(1975)}]{Polyakov1975}%
  \BibitemOpen
  \bibfield  {author} {\bibinfo {author} {\bibfnamefont {A.~M.}\ \bibnamefont
  {Polyakov}},\ }\href@noop {} {\bibfield  {journal} {\bibinfo  {journal}
  {Physics Letters B}\ }\textbf {\bibinfo {volume} {59}},\ \bibinfo {pages}
  {82} (\bibinfo {year} {1975})}\BibitemShut {NoStop}%
\bibitem [{\citenamefont {Belavin}\ \emph {et~al.}(1975)\citenamefont
  {Belavin}, \citenamefont {Polyakov}, \citenamefont {Schwartz},\ and\
  \citenamefont {Tyupkin}}]{Belavin1975}%
  \BibitemOpen
  \bibfield  {author} {\bibinfo {author} {\bibfnamefont {A.~A.}\ \bibnamefont
  {Belavin}}, \bibinfo {author} {\bibfnamefont {A.~M.}\ \bibnamefont
  {Polyakov}}, \bibinfo {author} {\bibfnamefont {A.~S.}\ \bibnamefont
  {Schwartz}}, \ and\ \bibinfo {author} {\bibfnamefont {Y.~S.}\ \bibnamefont
  {Tyupkin}},\ }\href@noop {} {\bibfield  {journal} {\bibinfo  {journal}
  {Physics Letters B}\ }\textbf {\bibinfo {volume} {59}},\ \bibinfo {pages}
  {85} (\bibinfo {year} {1975})}\BibitemShut {NoStop}%
\bibitem [{\citenamefont {{'t Hooft}}(1976)}]{tHooft1976}%
  \BibitemOpen
  \bibfield  {author} {\bibinfo {author} {\bibfnamefont {G.}~\bibnamefont {{'t
  Hooft}}},\ }\href {\doibase 10.1103/PhysRevLett.37.8} {\bibfield  {journal}
  {\bibinfo  {journal} {Physical Review Letters}\ }\textbf {\bibinfo {volume}
  {37}},\ \bibinfo {pages} {8} (\bibinfo {year} {1976})}\BibitemShut {NoStop}%
\bibitem [{\citenamefont {Kosterlitz}(1977)}]{Kosterlitz1977}%
  \BibitemOpen
  \bibfield  {author} {\bibinfo {author} {\bibfnamefont {J.~M.}\ \bibnamefont
  {Kosterlitz}},\ }\href@noop {} {\bibfield  {journal} {\bibinfo  {journal}
  {Journal of Physics C: Solid State Physics}\ }\textbf {\bibinfo {volume}
  {10}},\ \bibinfo {pages} {3753} (\bibinfo {year} {1977})}\BibitemShut
  {NoStop}%
\bibitem [{\citenamefont {Polyakov}(1977)}]{Polyakov1977}%
  \BibitemOpen
  \bibfield  {author} {\bibinfo {author} {\bibfnamefont {A.~M.}\ \bibnamefont
  {Polyakov}},\ }\href@noop {} {\bibfield  {journal} {\bibinfo  {journal}
  {Nuclear Physics B}\ }\textbf {\bibinfo {volume} {120}},\ \bibinfo {pages}
  {429} (\bibinfo {year} {1977})}\BibitemShut {NoStop}%
\bibitem [{\citenamefont {Polyakov}(1978)}]{Polyakov1978}%
  \BibitemOpen
  \bibfield  {author} {\bibinfo {author} {\bibfnamefont {A.~M.}\ \bibnamefont
  {Polyakov}},\ }\href@noop {} {\bibfield  {journal} {\bibinfo  {journal}
  {Physics Letters B}\ }\textbf {\bibinfo {volume} {72}},\ \bibinfo {pages}
  {477} (\bibinfo {year} {1978})}\BibitemShut {NoStop}%
\bibitem [{\citenamefont {Osterwalder}\ and\ \citenamefont
  {Seiler}(1978)}]{Osterwalder1978}%
  \BibitemOpen
  \bibfield  {author} {\bibinfo {author} {\bibfnamefont {K.}~\bibnamefont
  {Osterwalder}}\ and\ \bibinfo {author} {\bibfnamefont {E.}~\bibnamefont
  {Seiler}},\ }\href@noop {} {\bibfield  {journal} {\bibinfo  {journal} {Annals
  of Physics}\ }\textbf {\bibinfo {volume} {110}},\ \bibinfo {pages} {440}
  (\bibinfo {year} {1978})}\BibitemShut {NoStop}%
\bibitem [{\citenamefont {Guth}(1980)}]{Guth1980}%
  \BibitemOpen
  \bibfield  {author} {\bibinfo {author} {\bibfnamefont {A.~H.}\ \bibnamefont
  {Guth}},\ }\href {\doibase 10.1103/PhysRevD.21.2291} {\bibfield  {journal}
  {\bibinfo  {journal} {Physical Review D}\ }\textbf {\bibinfo {volume} {21}},\
  \bibinfo {pages} {2291} (\bibinfo {year} {1980})}\BibitemShut {NoStop}%
\bibitem [{\citenamefont {Lautrup}\ and\ \citenamefont
  {Nauenberg}(1980)}]{Lautrup1980}%
  \BibitemOpen
  \bibfield  {author} {\bibinfo {author} {\bibfnamefont {B.}~\bibnamefont
  {Lautrup}}\ and\ \bibinfo {author} {\bibfnamefont {M.}~\bibnamefont
  {Nauenberg}},\ }\href@noop {} {\bibfield  {journal} {\bibinfo  {journal}
  {Physics Letters B}\ }\textbf {\bibinfo {volume} {95}},\ \bibinfo {pages}
  {63} (\bibinfo {year} {1980})}\BibitemShut {NoStop}%
\bibitem [{\citenamefont {Svetitsky}\ and\ \citenamefont
  {Yaffe}(1982)}]{Yaffe1982}%
  \BibitemOpen
  \bibfield  {author} {\bibinfo {author} {\bibfnamefont {B.}~\bibnamefont
  {Svetitsky}}\ and\ \bibinfo {author} {\bibfnamefont {L.~G.}\ \bibnamefont
  {Yaffe}},\ }\href@noop {} {\bibfield  {journal} {\bibinfo  {journal} {Nuclear
  Physics B}\ }\textbf {\bibinfo {volume} {210}},\ \bibinfo {pages} {423}
  (\bibinfo {year} {1982})}\BibitemShut {NoStop}%
\bibitem [{\citenamefont {Nagaosa}(1993)}]{Nagaosa1993}%
  \BibitemOpen
  \bibfield  {author} {\bibinfo {author} {\bibfnamefont {N.}~\bibnamefont
  {Nagaosa}},\ }\href {\doibase 10.1103/PhysRevLett.71.4210} {\bibfield
  {journal} {\bibinfo  {journal} {Physical Review Letters}\ }\textbf {\bibinfo
  {volume} {71}},\ \bibinfo {pages} {4210} (\bibinfo {year}
  {1993})}\BibitemShut {NoStop}%
\bibitem [{\citenamefont {Sch{\"a}fer}\ and\ \citenamefont
  {Shuryak}(1998)}]{Shuryak1998}%
  \BibitemOpen
  \bibfield  {author} {\bibinfo {author} {\bibfnamefont {T.}~\bibnamefont
  {Sch{\"a}fer}}\ and\ \bibinfo {author} {\bibfnamefont {E.~V.}\ \bibnamefont
  {Shuryak}},\ }\href {\doibase 10.1103/RevModPhys.70.323} {\bibfield
  {journal} {\bibinfo  {journal} {Reviews of Modern Physics}\ }\textbf
  {\bibinfo {volume} {70}},\ \bibinfo {pages} {323} (\bibinfo {year}
  {1998})}\BibitemShut {NoStop}%
\bibitem [{\citenamefont {Kondo}(1998)}]{Kondo1998}%
  \BibitemOpen
  \bibfield  {author} {\bibinfo {author} {\bibfnamefont {K.-I.}\ \bibnamefont
  {Kondo}},\ }\href {\doibase 10.1103/PhysRevD.58.085013} {\bibfield  {journal}
  {\bibinfo  {journal} {Physical Review D}\ }\textbf {\bibinfo {volume} {58}},\
  \bibinfo {pages} {085013} (\bibinfo {year} {1998})}\BibitemShut {NoStop}%
\bibitem [{\citenamefont {Zheng}\ and\ \citenamefont
  {Sachdev}(1989)}]{zheng89}%
  \BibitemOpen
  \bibfield  {author} {\bibinfo {author} {\bibfnamefont {W.}~\bibnamefont
  {Zheng}}\ and\ \bibinfo {author} {\bibfnamefont {S.}~\bibnamefont
  {Sachdev}},\ }\href@noop {} {\bibfield  {journal} {\bibinfo  {journal}
  {Physical Review B}\ }\textbf {\bibinfo {volume} {40}},\ \bibinfo {pages}
  {2704} (\bibinfo {year} {1989})}\BibitemShut {NoStop}%
\bibitem [{\citenamefont {Read}\ and\ \citenamefont {Sachdev}(1990)}]{read90}%
  \BibitemOpen
  \bibfield  {author} {\bibinfo {author} {\bibfnamefont {N.}~\bibnamefont
  {Read}}\ and\ \bibinfo {author} {\bibfnamefont {S.}~\bibnamefont {Sachdev}},\
  }\href@noop {} {\bibfield  {journal} {\bibinfo  {journal} {Physical Review
  B}\ }\textbf {\bibinfo {volume} {42}},\ \bibinfo {pages} {4568} (\bibinfo
  {year} {1990})}\BibitemShut {NoStop}%
\bibitem [{\citenamefont {Kondo}(1999)}]{Kondo1999}%
  \BibitemOpen
  \bibfield  {author} {\bibinfo {author} {\bibfnamefont {K.-I.}\ \bibnamefont
  {Kondo}},\ }\href@noop {} {\bibfield  {journal} {\bibinfo  {journal} {Physics
  Letters B}\ }\textbf {\bibinfo {volume} {455}},\ \bibinfo {pages} {251}
  (\bibinfo {year} {1999})}\BibitemShut {NoStop}%
\bibitem [{\citenamefont {{{\rq}t Hooft}}\ and\ \citenamefont
  {Bruckmann}(2000)}]{tHooft2000}%
  \BibitemOpen
  \bibfield  {author} {\bibinfo {author} {\bibfnamefont {G.}~\bibnamefont
  {{{\rq}t Hooft}}}\ and\ \bibinfo {author} {\bibfnamefont {F.}~\bibnamefont
  {Bruckmann}},\ }\href@noop {} {\  (\bibinfo {year} {2000})},\ \bibinfo {note}
  {arXiv:hep-th/0010225}\BibitemShut {NoStop}%
\bibitem [{\citenamefont {Nagaosa}\ and\ \citenamefont
  {Lee}(2000)}]{Nagaosa2000}%
  \BibitemOpen
  \bibfield  {author} {\bibinfo {author} {\bibfnamefont {N.}~\bibnamefont
  {Nagaosa}}\ and\ \bibinfo {author} {\bibfnamefont {P.~A.}\ \bibnamefont
  {Lee}},\ }\href@noop {} {\bibfield  {journal} {\bibinfo  {journal} {Physical
  Review B}\ }\textbf {\bibinfo {volume} {61}},\ \bibinfo {pages} {9166}
  (\bibinfo {year} {2000})}\BibitemShut {NoStop}%
\bibitem [{\citenamefont {Yoshida}(2001)}]{Yoshida2001}%
  \BibitemOpen
  \bibfield  {author} {\bibinfo {author} {\bibfnamefont {K.}~\bibnamefont
  {Yoshida}},\ }\href@noop {} {\bibfield  {journal} {\bibinfo  {journal}
  {Journal of High Energy Physics}\ }\textbf {\bibinfo {volume} {04}},\
  \bibinfo {pages} {030} (\bibinfo {year} {2001})}\BibitemShut {NoStop}%
\bibitem [{\citenamefont {Sachdev}\ and\ \citenamefont
  {Park}(2002)}]{sachdev02e}%
  \BibitemOpen
  \bibfield  {author} {\bibinfo {author} {\bibfnamefont {S.}~\bibnamefont
  {Sachdev}}\ and\ \bibinfo {author} {\bibfnamefont {K.}~\bibnamefont {Park}},\
  }\href@noop {} {\bibfield  {journal} {\bibinfo  {journal} {Annals of
  Physics}\ }\textbf {\bibinfo {volume} {298}},\ \bibinfo {pages} {58}
  (\bibinfo {year} {2002})}\BibitemShut {NoStop}%
\bibitem [{\citenamefont {Kleinert}\ \emph {et~al.}(2002)\citenamefont
  {Kleinert}, \citenamefont {Nogueira},\ and\ \citenamefont
  {Sudb{\o}}}]{Sudbo2002}%
  \BibitemOpen
  \bibfield  {author} {\bibinfo {author} {\bibfnamefont {H.}~\bibnamefont
  {Kleinert}}, \bibinfo {author} {\bibfnamefont {F.~S.}\ \bibnamefont
  {Nogueira}}, \ and\ \bibinfo {author} {\bibfnamefont {A.}~\bibnamefont
  {Sudb{\o}}},\ }\href {\doibase 10.1103/PhysRevLett.88.232001} {\bibfield
  {journal} {\bibinfo  {journal} {Physical Review Letters}\ }\textbf {\bibinfo
  {volume} {88}},\ \bibinfo {pages} {232001} (\bibinfo {year}
  {2002})}\BibitemShut {NoStop}%
\bibitem [{\citenamefont {Kragset}\ \emph {et~al.}(2004)\citenamefont
  {Kragset}, \citenamefont {Sudb{\o}},\ and\ \citenamefont
  {Nogueira}}]{Sudbo2004}%
  \BibitemOpen
  \bibfield  {author} {\bibinfo {author} {\bibfnamefont {S.}~\bibnamefont
  {Kragset}}, \bibinfo {author} {\bibfnamefont {A.}~\bibnamefont {Sudb{\o}}}, \
  and\ \bibinfo {author} {\bibfnamefont {F.~S.}\ \bibnamefont {Nogueira}},\
  }\href {\doibase 10.1103/PhysRevLett.92.186403} {\bibfield  {journal}
  {\bibinfo  {journal} {Physical Review Letters}\ }\textbf {\bibinfo {volume}
  {92}},\ \bibinfo {pages} {186403} (\bibinfo {year} {2004})}\BibitemShut
  {NoStop}%
\bibitem [{\citenamefont {Herbut}\ and\ \citenamefont
  {Seradjeh}(2003)}]{Herbut2003}%
  \BibitemOpen
  \bibfield  {author} {\bibinfo {author} {\bibfnamefont {I.~F.}\ \bibnamefont
  {Herbut}}\ and\ \bibinfo {author} {\bibfnamefont {B.~H.}\ \bibnamefont
  {Seradjeh}},\ }\href {\doibase 10.1103/PhysRevLett.91.171601} {\bibfield
  {journal} {\bibinfo  {journal} {Physical Review Letters}\ }\textbf {\bibinfo
  {volume} {91}},\ \bibinfo {pages} {171601} (\bibinfo {year}
  {2003})}\BibitemShut {NoStop}%
\bibitem [{\citenamefont {Case}\ \emph {et~al.}(2004)\citenamefont {Case},
  \citenamefont {Seradjeh},\ and\ \citenamefont {Herbut}}]{Herbut2004}%
  \BibitemOpen
  \bibfield  {author} {\bibinfo {author} {\bibfnamefont {M.~J.}\ \bibnamefont
  {Case}}, \bibinfo {author} {\bibfnamefont {B.~H.}\ \bibnamefont {Seradjeh}},
  \ and\ \bibinfo {author} {\bibfnamefont {I.~F.}\ \bibnamefont {Herbut}},\
  }\href@noop {} {\bibfield  {journal} {\bibinfo  {journal} {Nuclear Physics
  B}\ }\textbf {\bibinfo {volume} {676}},\ \bibinfo {pages} {572} (\bibinfo
  {year} {2004})}\BibitemShut {NoStop}%
\bibitem [{\citenamefont {Herbut}\ \emph {et~al.}(2003)\citenamefont {Herbut},
  \citenamefont {Seradjeh}, \citenamefont {Sachdev},\ and\ \citenamefont
  {Murthy}}]{herbut03}%
  \BibitemOpen
  \bibfield  {author} {\bibinfo {author} {\bibfnamefont {I.~F.}\ \bibnamefont
  {Herbut}}, \bibinfo {author} {\bibfnamefont {B.~H.}\ \bibnamefont
  {Seradjeh}}, \bibinfo {author} {\bibfnamefont {S.}~\bibnamefont {Sachdev}}, \
  and\ \bibinfo {author} {\bibfnamefont {G.}~\bibnamefont {Murthy}},\
  }\href@noop {} {\bibfield  {journal} {\bibinfo  {journal} {Physical Review
  B}\ }\textbf {\bibinfo {volume} {68}},\ \bibinfo {pages} {195110} (\bibinfo
  {year} {2003})}\BibitemShut {NoStop}%
\bibitem [{\citenamefont {Hermele}\ \emph {et~al.}(2004)\citenamefont
  {Hermele}, \citenamefont {Senthil}, \citenamefont {Fisher}, \citenamefont
  {Lee}, \citenamefont {Nagaosa},\ and\ \citenamefont {Wen}}]{Hermele2004}%
  \BibitemOpen
  \bibfield  {author} {\bibinfo {author} {\bibfnamefont {M.}~\bibnamefont
  {Hermele}}, \bibinfo {author} {\bibfnamefont {T.}~\bibnamefont {Senthil}},
  \bibinfo {author} {\bibfnamefont {M.~P.~A.}\ \bibnamefont {Fisher}}, \bibinfo
  {author} {\bibfnamefont {P.~A.}\ \bibnamefont {Lee}}, \bibinfo {author}
  {\bibfnamefont {N.}~\bibnamefont {Nagaosa}}, \ and\ \bibinfo {author}
  {\bibfnamefont {X.~G.}\ \bibnamefont {Wen}},\ }\href@noop {} {\bibfield
  {journal} {\bibinfo  {journal} {Physical Review B}\ }\textbf {\bibinfo
  {volume} {70}},\ \bibinfo {pages} {214437} (\bibinfo {year}
  {2004})}\BibitemShut {NoStop}%
\bibitem [{\citenamefont {B{\o}rkje}\ \emph {et~al.}(2005)\citenamefont
  {B{\o}rkje}, \citenamefont {Kragset},\ and\ \citenamefont
  {Sudb{\o}}}]{Sudbo2005}%
  \BibitemOpen
  \bibfield  {author} {\bibinfo {author} {\bibfnamefont {K.}~\bibnamefont
  {B{\o}rkje}}, \bibinfo {author} {\bibfnamefont {S.}~\bibnamefont {Kragset}},
  \ and\ \bibinfo {author} {\bibfnamefont {A.}~\bibnamefont {Sudb{\o}}},\
  }\href {\doibase 10.1103/PhysRevB.71.085112} {\bibfield  {journal} {\bibinfo
  {journal} {Physical Review B}\ }\textbf {\bibinfo {volume} {71}},\ \bibinfo
  {pages} {085112} (\bibinfo {year} {2005})}\BibitemShut {NoStop}%
\bibitem [{\citenamefont {Engelhardt}(2005)}]{Engelhardt2005}%
  \BibitemOpen
  \bibfield  {author} {\bibinfo {author} {\bibfnamefont {M.}~\bibnamefont
  {Engelhardt}},\ }\href@noop {} {\bibfield  {journal} {\bibinfo  {journal}
  {Nuclear Physics B - Proceedings Supplements}\ }\textbf {\bibinfo {volume}
  {140}},\ \bibinfo {pages} {92} (\bibinfo {year} {2005})}\BibitemShut
  {NoStop}%
\bibitem [{\citenamefont {Wang}(2005)}]{Wang2005a}%
  \BibitemOpen
  \bibfield  {author} {\bibinfo {author} {\bibfnamefont {Z.}~\bibnamefont
  {Wang}},\ }\href {\doibase 10.1103/PhysRevLett.94.176804} {\bibfield
  {journal} {\bibinfo  {journal} {Physical Review Letters}\ }\textbf {\bibinfo
  {volume} {94}},\ \bibinfo {pages} {176804} (\bibinfo {year}
  {2005})}\BibitemShut {NoStop}%
\bibitem [{\citenamefont {Nogueira}\ and\ \citenamefont
  {Kleinert}(2005)}]{Kleinert2005}%
  \BibitemOpen
  \bibfield  {author} {\bibinfo {author} {\bibfnamefont {F.~S.}\ \bibnamefont
  {Nogueira}}\ and\ \bibinfo {author} {\bibfnamefont {H.}~\bibnamefont
  {Kleinert}},\ }\href {\doibase 10.1103/PhysRevLett.95.176406} {\bibfield
  {journal} {\bibinfo  {journal} {Physical Review Letters}\ }\textbf {\bibinfo
  {volume} {95}},\ \bibinfo {pages} {176406} (\bibinfo {year}
  {2005})}\BibitemShut {NoStop}%
\bibitem [{\citenamefont {Nogueira}\ and\ \citenamefont
  {Kleinert}(2008)}]{Nogueira2008}%
  \BibitemOpen
  \bibfield  {author} {\bibinfo {author} {\bibfnamefont {F.~S.}\ \bibnamefont
  {Nogueira}}\ and\ \bibinfo {author} {\bibfnamefont {H.}~\bibnamefont
  {Kleinert}},\ }\href {\doibase 10.1103/PhysRevB.77.045107} {\bibfield
  {journal} {\bibinfo  {journal} {Physical Review B}\ }\textbf {\bibinfo
  {volume} {77}},\ \bibinfo {pages} {045107} (\bibinfo {year}
  {2008})}\BibitemShut {NoStop}%
\bibitem [{\citenamefont {Yamamoto}(2008)}]{Yamamoto2008}%
  \BibitemOpen
  \bibfield  {author} {\bibinfo {author} {\bibfnamefont {N.}~\bibnamefont
  {Yamamoto}},\ }\href@noop {} {\bibfield  {journal} {\bibinfo  {journal}
  {Journal of High Energy Physics}\ }\textbf {\bibinfo {volume} {12}},\
  \bibinfo {pages} {060} (\bibinfo {year} {2008})}\BibitemShut {NoStop}%
\bibitem [{\citenamefont {Ogilvie}(2012)}]{Oglivie2012}%
  \BibitemOpen
  \bibfield  {author} {\bibinfo {author} {\bibfnamefont {M.~C.}\ \bibnamefont
  {Ogilvie}},\ }\href@noop {} {\bibfield  {journal} {\bibinfo  {journal}
  {Journal of Physics A: Mathematical and Theoretical}\ }\textbf {\bibinfo
  {volume} {45}},\ \bibinfo {pages} {483001} (\bibinfo {year}
  {2012})}\BibitemShut {NoStop}%
\bibitem [{\citenamefont {Anber}(2014)}]{Anber2014}%
  \BibitemOpen
  \bibfield  {author} {\bibinfo {author} {\bibfnamefont {M.~M.}\ \bibnamefont
  {Anber}},\ }\href@noop {} {\bibfield  {journal} {\bibinfo  {journal} {Annals
  of Physics}\ }\textbf {\bibinfo {volume} {341}},\ \bibinfo {pages} {21}
  (\bibinfo {year} {2014})}\BibitemShut {NoStop}%
\bibitem [{\citenamefont {Heydeman}\ \emph {et~al.}(2023)\citenamefont
  {Heydeman}, \citenamefont {Jepsen}, \citenamefont {Ji},\ and\ \citenamefont
  {Yarom}}]{Heydeman2023}%
  \BibitemOpen
  \bibfield  {author} {\bibinfo {author} {\bibfnamefont {M.}~\bibnamefont
  {Heydeman}}, \bibinfo {author} {\bibfnamefont {C.~B.}\ \bibnamefont
  {Jepsen}}, \bibinfo {author} {\bibfnamefont {Z.}~\bibnamefont {Ji}}, \ and\
  \bibinfo {author} {\bibfnamefont {A.}~\bibnamefont {Yarom}},\ }\href@noop {}
  {\bibfield  {journal} {\bibinfo  {journal} {Journal of High Energy Physics}\
  }\textbf {\bibinfo {volume} {2023}},\ \bibinfo {pages} {119} (\bibinfo {year}
  {2023})}\BibitemShut {NoStop}%
\bibitem [{\citenamefont {Wang}\ \emph {et~al.}(2001)\citenamefont {Wang},
  \citenamefont {Xu}, \citenamefont {Kakeshita}, \citenamefont {Uchida},
  \citenamefont {Ono}, \citenamefont {Ando},\ and\ \citenamefont
  {Ong}}]{Ong2001}%
  \BibitemOpen
  \bibfield  {author} {\bibinfo {author} {\bibfnamefont {Y.}~\bibnamefont
  {Wang}}, \bibinfo {author} {\bibfnamefont {Z.~A.}\ \bibnamefont {Xu}},
  \bibinfo {author} {\bibfnamefont {T.}~\bibnamefont {Kakeshita}}, \bibinfo
  {author} {\bibfnamefont {S.}~\bibnamefont {Uchida}}, \bibinfo {author}
  {\bibfnamefont {S.}~\bibnamefont {Ono}}, \bibinfo {author} {\bibfnamefont
  {Y.}~\bibnamefont {Ando}}, \ and\ \bibinfo {author} {\bibfnamefont {N.~P.}\
  \bibnamefont {Ong}},\ }\href@noop {} {\bibfield  {journal} {\bibinfo
  {journal} {Physical Review B}\ }\textbf {\bibinfo {volume} {64}},\ \bibinfo
  {pages} {224519} (\bibinfo {year} {2001})}\BibitemShut {NoStop}%
\bibitem [{\citenamefont {Wang}\ \emph {et~al.}(2006)\citenamefont {Wang},
  \citenamefont {Li},\ and\ \citenamefont {Ong}}]{Wang2006}%
  \BibitemOpen
  \bibfield  {author} {\bibinfo {author} {\bibfnamefont {Y.}~\bibnamefont
  {Wang}}, \bibinfo {author} {\bibfnamefont {L.}~\bibnamefont {Li}}, \ and\
  \bibinfo {author} {\bibfnamefont {N.~P.}\ \bibnamefont {Ong}},\ }\href@noop
  {} {\bibfield  {journal} {\bibinfo  {journal} {Physical Review B}\ }\textbf
  {\bibinfo {volume} {73}},\ \bibinfo {eid} {024510} (\bibinfo {year}
  {2006})}\BibitemShut {NoStop}%
\bibitem [{\citenamefont {Li}\ \emph {et~al.}(2010)\citenamefont {Li},
  \citenamefont {Wang}, \citenamefont {Komiya}, \citenamefont {Ono},
  \citenamefont {Ando}, \citenamefont {Gu},\ and\ \citenamefont
  {Ong}}]{Li2010}%
  \BibitemOpen
  \bibfield  {author} {\bibinfo {author} {\bibfnamefont {L.}~\bibnamefont
  {Li}}, \bibinfo {author} {\bibfnamefont {Y.}~\bibnamefont {Wang}}, \bibinfo
  {author} {\bibfnamefont {S.}~\bibnamefont {Komiya}}, \bibinfo {author}
  {\bibfnamefont {S.}~\bibnamefont {Ono}}, \bibinfo {author} {\bibfnamefont
  {Y.}~\bibnamefont {Ando}}, \bibinfo {author} {\bibfnamefont {G.~D.}\
  \bibnamefont {Gu}}, \ and\ \bibinfo {author} {\bibfnamefont {N.~P.}\
  \bibnamefont {Ong}},\ }\href@noop {} {\bibfield  {journal} {\bibinfo
  {journal} {Physical Review B}\ }\textbf {\bibinfo {volume} {81}},\ \bibinfo
  {pages} {054510} (\bibinfo {year} {2010})}\BibitemShut {NoStop}%
\bibitem [{\citenamefont {Nikoli{\'c}}(2020{\natexlab{a}})}]{Nikolic2019}%
  \BibitemOpen
  \bibfield  {author} {\bibinfo {author} {\bibfnamefont {P.}~\bibnamefont
  {Nikoli{\'c}}},\ }\href@noop {} {\bibfield  {journal} {\bibinfo  {journal}
  {Physical Review B}\ }\textbf {\bibinfo {volume} {101}},\ \bibinfo {pages}
  {115144} (\bibinfo {year} {2020}{\natexlab{a}})}\BibitemShut {NoStop}%
\bibitem [{\citenamefont {Fisher}\ \emph {et~al.}(1989)\citenamefont {Fisher},
  \citenamefont {Weichman}, \citenamefont {Grinstein},\ and\ \citenamefont
  {Fisher}}]{Fisher1989a}%
  \BibitemOpen
  \bibfield  {author} {\bibinfo {author} {\bibfnamefont {M.~P.~A.}\
  \bibnamefont {Fisher}}, \bibinfo {author} {\bibfnamefont {P.~B.}\
  \bibnamefont {Weichman}}, \bibinfo {author} {\bibfnamefont {G.}~\bibnamefont
  {Grinstein}}, \ and\ \bibinfo {author} {\bibfnamefont {D.~S.}\ \bibnamefont
  {Fisher}},\ }\href@noop {} {\bibfield  {journal} {\bibinfo  {journal}
  {Physical Review B}\ }\textbf {\bibinfo {volume} {40}},\ \bibinfo {pages}
  {546} (\bibinfo {year} {1989})}\BibitemShut {NoStop}%
\bibitem [{\citenamefont {Fisher}(1990)}]{Fisher1990}%
  \BibitemOpen
  \bibfield  {author} {\bibinfo {author} {\bibfnamefont {M.~P.~A.}\
  \bibnamefont {Fisher}},\ }\href@noop {} {\bibfield  {journal} {\bibinfo
  {journal} {Physical Review Letters}\ }\textbf {\bibinfo {volume} {65}},\
  \bibinfo {pages} {923} (\bibinfo {year} {1990})}\BibitemShut {NoStop}%
\bibitem [{\citenamefont {Elitzur}(1975)}]{Elitzur1975}%
  \BibitemOpen
  \bibfield  {author} {\bibinfo {author} {\bibfnamefont {S.}~\bibnamefont
  {Elitzur}},\ }\href@noop {} {\bibfield  {journal} {\bibinfo  {journal}
  {Physical Review D}\ }\textbf {\bibinfo {volume} {12}},\ \bibinfo {pages}
  {3978} (\bibinfo {year} {1975})}\BibitemShut {NoStop}%
\bibitem [{\citenamefont {Fradkin}\ and\ \citenamefont
  {Shenker}(1979)}]{fradkin79a}%
  \BibitemOpen
  \bibfield  {author} {\bibinfo {author} {\bibfnamefont {E.}~\bibnamefont
  {Fradkin}}\ and\ \bibinfo {author} {\bibfnamefont {S.~H.}\ \bibnamefont
  {Shenker}},\ }\href@noop {} {\bibfield  {journal} {\bibinfo  {journal}
  {Physical Review D}\ }\textbf {\bibinfo {volume} {19}},\ \bibinfo {pages}
  {3682} (\bibinfo {year} {1979})}\BibitemShut {NoStop}%
\bibitem [{\citenamefont {Caillol}\ and\ \citenamefont
  {Levesque}(1986)}]{Levesque1986}%
  \BibitemOpen
  \bibfield  {author} {\bibinfo {author} {\bibfnamefont {J.~M.}\ \bibnamefont
  {Caillol}}\ and\ \bibinfo {author} {\bibfnamefont {D.}~\bibnamefont
  {Levesque}},\ }\href {\doibase 10.1103/PhysRevB.33.499} {\bibfield  {journal}
  {\bibinfo  {journal} {Physical Review B}\ }\textbf {\bibinfo {volume} {33}},\
  \bibinfo {pages} {499} (\bibinfo {year} {1986})}\BibitemShut {NoStop}%
\bibitem [{\citenamefont {Minnhagen}(1987)}]{Minnhagen1987}%
  \BibitemOpen
  \bibfield  {author} {\bibinfo {author} {\bibfnamefont {P.}~\bibnamefont
  {Minnhagen}},\ }\href {\doibase 10.1103/RevModPhys.59.1001} {\bibfield
  {journal} {\bibinfo  {journal} {Reviews of Modern Physics}\ }\textbf
  {\bibinfo {volume} {59}},\ \bibinfo {pages} {1001} (\bibinfo {year}
  {1987})}\BibitemShut {NoStop}%
\bibitem [{\citenamefont {Levin}\ \emph {et~al.}(1994)\citenamefont {Levin},
  \citenamefont {jun Li},\ and\ \citenamefont {Fisher}}]{Fisher1994}%
  \BibitemOpen
  \bibfield  {author} {\bibinfo {author} {\bibfnamefont {Y.}~\bibnamefont
  {Levin}}, \bibinfo {author} {\bibfnamefont {X.}~\bibnamefont {jun Li}}, \
  and\ \bibinfo {author} {\bibfnamefont {M.~E.}\ \bibnamefont {Fisher}},\
  }\href {\doibase 10.1103/PhysRevLett.73.2716} {\bibfield  {journal} {\bibinfo
   {journal} {Physical Review Letters}\ }\textbf {\bibinfo {volume} {73}},\
  \bibinfo {pages} {2716} (\bibinfo {year} {1994})}\BibitemShut {NoStop}%
\bibitem [{\citenamefont {Fisher}(1994)}]{Fisher1994b}%
  \BibitemOpen
  \bibfield  {author} {\bibinfo {author} {\bibfnamefont {M.~E.}\ \bibnamefont
  {Fisher}},\ }\href@noop {} {\bibfield  {journal} {\bibinfo  {journal}
  {Journal of Statistical Physics}\ }\textbf {\bibinfo {volume} {75}},\
  \bibinfo {pages} {1} (\bibinfo {year} {1994})}\BibitemShut {NoStop}%
\bibitem [{\citenamefont {{Gama Goicochea}}\ and\ \citenamefont
  {Nussinov}(2023)}]{Nussinov2023}%
  \BibitemOpen
  \bibfield  {author} {\bibinfo {author} {\bibfnamefont {A.}~\bibnamefont
  {{Gama Goicochea}}}\ and\ \bibinfo {author} {\bibfnamefont {Z.}~\bibnamefont
  {Nussinov}},\ }\href {\doibase 10.1103/PhysRevE.107.014104} {\bibfield
  {journal} {\bibinfo  {journal} {Physical Review E}\ }\textbf {\bibinfo
  {volume} {107}},\ \bibinfo {pages} {014104} (\bibinfo {year}
  {2023})}\BibitemShut {NoStop}%
\bibitem [{Note1()}]{Note1}%
  \BibitemOpen
  \bibinfo {note} {Sufficiently high order terms in this expansion create
  infra-red divergence, but this is an artifact of the expansion. The exact
  integral is convergent and approximated by its well-behaved lowest order
  term.}\BibitemShut {Stop}%
\bibitem [{\citenamefont {Nikoli{\'c}}\ and\ \citenamefont
  {Tesanovi{\'c}}(2011)}]{Nikolic2010b}%
  \BibitemOpen
  \bibfield  {author} {\bibinfo {author} {\bibfnamefont {P.}~\bibnamefont
  {Nikoli{\'c}}}\ and\ \bibinfo {author} {\bibfnamefont {Z.}~\bibnamefont
  {Tesanovi{\'c}}},\ }\href@noop {} {\bibfield  {journal} {\bibinfo  {journal}
  {Physical Review B}\ }\textbf {\bibinfo {volume} {83}},\ \bibinfo {pages}
  {064501} (\bibinfo {year} {2011})}\BibitemShut {NoStop}%
\bibitem [{\citenamefont {Moon}\ \emph {et~al.}(2007)\citenamefont {Moon},
  \citenamefont {Nikoli{\'c}},\ and\ \citenamefont {Sachdev}}]{moon:230403}%
  \BibitemOpen
  \bibfield  {author} {\bibinfo {author} {\bibfnamefont {E.~G.}\ \bibnamefont
  {Moon}}, \bibinfo {author} {\bibfnamefont {P.}~\bibnamefont {Nikoli{\'c}}}, \
  and\ \bibinfo {author} {\bibfnamefont {S.}~\bibnamefont {Sachdev}},\ }\href
  {\doibase 10.1103/PhysRevLett.99.230403} {\bibfield  {journal} {\bibinfo
  {journal} {Physical Review Letters}\ }\textbf {\bibinfo {volume} {99}},\
  \bibinfo {eid} {230403} (\bibinfo {year} {2007})}\BibitemShut {NoStop}%
\bibitem [{\citenamefont {Nikoli{\'c}}(2011)}]{Nikolic2010}%
  \BibitemOpen
  \bibfield  {author} {\bibinfo {author} {\bibfnamefont {P.}~\bibnamefont
  {Nikoli{\'c}}},\ }\href@noop {} {\bibfield  {journal} {\bibinfo  {journal}
  {Physical Review B}\ }\textbf {\bibinfo {volume} {83}},\ \bibinfo {pages}
  {064523} (\bibinfo {year} {2011})}\BibitemShut {NoStop}%
\bibitem [{\citenamefont {Corson}\ \emph {et~al.}(1999)\citenamefont {Corson},
  \citenamefont {Mallozzi}, \citenamefont {Orenstein}, \citenamefont
  {Eckstein},\ and\ \citenamefont {Bozovic}}]{Corson1999}%
  \BibitemOpen
  \bibfield  {author} {\bibinfo {author} {\bibfnamefont {J.}~\bibnamefont
  {Corson}}, \bibinfo {author} {\bibfnamefont {R.}~\bibnamefont {Mallozzi}},
  \bibinfo {author} {\bibfnamefont {J.}~\bibnamefont {Orenstein}}, \bibinfo
  {author} {\bibfnamefont {J.~N.}\ \bibnamefont {Eckstein}}, \ and\ \bibinfo
  {author} {\bibfnamefont {I.}~\bibnamefont {Bozovic}},\ }\href@noop {}
  {\bibfield  {journal} {\bibinfo  {journal} {Nature}\ }\textbf {\bibinfo
  {volume} {398}},\ \bibinfo {pages} {221} (\bibinfo {year}
  {1999})}\BibitemShut {NoStop}%
\bibitem [{\citenamefont {Vershinin}\ \emph {et~al.}(2004)\citenamefont
  {Vershinin}, \citenamefont {Misra}, \citenamefont {Ono}, \citenamefont {Abe},
  \citenamefont {Ando},\ and\ \citenamefont {Yazdani}}]{Yazdani2004}%
  \BibitemOpen
  \bibfield  {author} {\bibinfo {author} {\bibfnamefont {M.}~\bibnamefont
  {Vershinin}}, \bibinfo {author} {\bibfnamefont {S.}~\bibnamefont {Misra}},
  \bibinfo {author} {\bibfnamefont {S.}~\bibnamefont {Ono}}, \bibinfo {author}
  {\bibfnamefont {Y.}~\bibnamefont {Abe}}, \bibinfo {author} {\bibfnamefont
  {Y.}~\bibnamefont {Ando}}, \ and\ \bibinfo {author} {\bibfnamefont
  {A.}~\bibnamefont {Yazdani}},\ }\href@noop {} {\bibfield  {journal} {\bibinfo
   {journal} {Science}\ }\textbf {\bibinfo {volume} {303}},\ \bibinfo {pages}
  {1995} (\bibinfo {year} {2004})}\BibitemShut {NoStop}%
\bibitem [{\citenamefont {Fang}\ \emph {et~al.}(2004)\citenamefont {Fang},
  \citenamefont {Howald}, \citenamefont {Kaneko}, \citenamefont {Greven},\ and\
  \citenamefont {Kapitulnik}}]{Fang2004}%
  \BibitemOpen
  \bibfield  {author} {\bibinfo {author} {\bibfnamefont {A.}~\bibnamefont
  {Fang}}, \bibinfo {author} {\bibfnamefont {C.}~\bibnamefont {Howald}},
  \bibinfo {author} {\bibfnamefont {N.}~\bibnamefont {Kaneko}}, \bibinfo
  {author} {\bibfnamefont {M.}~\bibnamefont {Greven}}, \ and\ \bibinfo {author}
  {\bibfnamefont {A.}~\bibnamefont {Kapitulnik}},\ }\href@noop {} {\bibfield
  {journal} {\bibinfo  {journal} {Physical Review B}\ }\textbf {\bibinfo
  {volume} {70}},\ \bibinfo {pages} {214514} (\bibinfo {year}
  {2004})}\BibitemShut {NoStop}%
\bibitem [{\citenamefont {Steiner}\ \emph {et~al.}(2005)\citenamefont
  {Steiner}, \citenamefont {Boebinger},\ and\ \citenamefont
  {Kapitulnik}}]{Steiner2005}%
  \BibitemOpen
  \bibfield  {author} {\bibinfo {author} {\bibfnamefont {M.~A.}\ \bibnamefont
  {Steiner}}, \bibinfo {author} {\bibfnamefont {G.}~\bibnamefont {Boebinger}},
  \ and\ \bibinfo {author} {\bibfnamefont {A.}~\bibnamefont {Kapitulnik}},\
  }\href@noop {} {\bibfield  {journal} {\bibinfo  {journal} {Physical Review
  Letters}\ }\textbf {\bibinfo {volume} {94}},\ \bibinfo {pages} {107008}
  (\bibinfo {year} {2005})}\BibitemShut {NoStop}%
\bibitem [{\citenamefont {Valla}\ \emph {et~al.}(2006)\citenamefont {Valla},
  \citenamefont {Fedorov}, \citenamefont {Lee}, \citenamefont {Davis},\ and\
  \citenamefont {Gu}}]{Valla2006}%
  \BibitemOpen
  \bibfield  {author} {\bibinfo {author} {\bibfnamefont {T.}~\bibnamefont
  {Valla}}, \bibinfo {author} {\bibfnamefont {A.~V.}\ \bibnamefont {Fedorov}},
  \bibinfo {author} {\bibfnamefont {J.}~\bibnamefont {Lee}}, \bibinfo {author}
  {\bibfnamefont {J.~C.}\ \bibnamefont {Davis}}, \ and\ \bibinfo {author}
  {\bibfnamefont {G.~D.}\ \bibnamefont {Gu}},\ }\href@noop {} {\bibfield
  {journal} {\bibinfo  {journal} {Science}\ }\textbf {\bibinfo {volume}
  {314}},\ \bibinfo {pages} {1914} (\bibinfo {year} {2006})}\BibitemShut
  {NoStop}%
\bibitem [{\citenamefont {Crane}\ \emph {et~al.}(2007)\citenamefont {Crane},
  \citenamefont {Armitage}, \citenamefont {Johansson}, \citenamefont
  {Sambandamurthy}, \citenamefont {Shahar},\ and\ \citenamefont
  {Gruner}}]{Armitage2007}%
  \BibitemOpen
  \bibfield  {author} {\bibinfo {author} {\bibfnamefont {R.~W.}\ \bibnamefont
  {Crane}}, \bibinfo {author} {\bibfnamefont {N.~P.}\ \bibnamefont {Armitage}},
  \bibinfo {author} {\bibfnamefont {A.}~\bibnamefont {Johansson}}, \bibinfo
  {author} {\bibfnamefont {G.}~\bibnamefont {Sambandamurthy}}, \bibinfo
  {author} {\bibfnamefont {D.}~\bibnamefont {Shahar}}, \ and\ \bibinfo {author}
  {\bibfnamefont {G.}~\bibnamefont {Gruner}},\ }\href@noop {} {\bibfield
  {journal} {\bibinfo  {journal} {Physical Review B}\ }\textbf {\bibinfo
  {volume} {75}},\ \bibinfo {pages} {184530} (\bibinfo {year}
  {2007})}\BibitemShut {NoStop}%
\bibitem [{\citenamefont {Boyer}\ \emph {et~al.}(2007)\citenamefont {Boyer},
  \citenamefont {Wise}, \citenamefont {Chatterjee}, \citenamefont {Yi},
  \citenamefont {Kondo}, \citenamefont {Takeuchi}, \citenamefont {Ikuta},\ and\
  \citenamefont {Hudson}}]{Hudson2007}%
  \BibitemOpen
  \bibfield  {author} {\bibinfo {author} {\bibfnamefont {M.~C.}\ \bibnamefont
  {Boyer}}, \bibinfo {author} {\bibfnamefont {W.~D.}\ \bibnamefont {Wise}},
  \bibinfo {author} {\bibfnamefont {K.}~\bibnamefont {Chatterjee}}, \bibinfo
  {author} {\bibfnamefont {M.}~\bibnamefont {Yi}}, \bibinfo {author}
  {\bibfnamefont {T.}~\bibnamefont {Kondo}}, \bibinfo {author} {\bibfnamefont
  {T.}~\bibnamefont {Takeuchi}}, \bibinfo {author} {\bibfnamefont
  {H.}~\bibnamefont {Ikuta}}, \ and\ \bibinfo {author} {\bibfnamefont {E.~W.}\
  \bibnamefont {Hudson}},\ }\href@noop {} {\bibfield  {journal} {\bibinfo
  {journal} {Nature Physics}\ }\textbf {\bibinfo {volume} {3}},\ \bibinfo
  {pages} {802} (\bibinfo {year} {2007})}\BibitemShut {NoStop}%
\bibitem [{\citenamefont {Kohsaka}\ \emph {et~al.}(2007)\citenamefont
  {Kohsaka}, \citenamefont {Taylor}, \citenamefont {Fujita}, \citenamefont
  {Schmidt}, \citenamefont {Lupien}, \citenamefont {Hanaguri}, \citenamefont
  {Azuma}, \citenamefont {Takano}, \citenamefont {Eisaki}, \citenamefont
  {Takagi}, \citenamefont {Uchida},\ and\ \citenamefont {Davis}}]{Kohsaka2007}%
  \BibitemOpen
  \bibfield  {author} {\bibinfo {author} {\bibfnamefont {Y.}~\bibnamefont
  {Kohsaka}}, \bibinfo {author} {\bibfnamefont {C.}~\bibnamefont {Taylor}},
  \bibinfo {author} {\bibfnamefont {K.}~\bibnamefont {Fujita}}, \bibinfo
  {author} {\bibfnamefont {A.}~\bibnamefont {Schmidt}}, \bibinfo {author}
  {\bibfnamefont {C.}~\bibnamefont {Lupien}}, \bibinfo {author} {\bibfnamefont
  {T.}~\bibnamefont {Hanaguri}}, \bibinfo {author} {\bibfnamefont
  {M.}~\bibnamefont {Azuma}}, \bibinfo {author} {\bibfnamefont
  {M.}~\bibnamefont {Takano}}, \bibinfo {author} {\bibfnamefont
  {H.}~\bibnamefont {Eisaki}}, \bibinfo {author} {\bibfnamefont
  {H.}~\bibnamefont {Takagi}}, \bibinfo {author} {\bibfnamefont
  {S.}~\bibnamefont {Uchida}}, \ and\ \bibinfo {author} {\bibfnamefont {J.~C.}\
  \bibnamefont {Davis}},\ }\href@noop {} {\bibfield  {journal} {\bibinfo
  {journal} {Science}\ }\textbf {\bibinfo {volume} {315}},\ \bibinfo {pages}
  {1380} (\bibinfo {year} {2007})}\BibitemShut {NoStop}%
\bibitem [{\citenamefont {Gomes}\ \emph {et~al.}(2007)\citenamefont {Gomes},
  \citenamefont {Pasupathy}, \citenamefont {Pushp}, \citenamefont {Ono},
  \citenamefont {Ando},\ and\ \citenamefont {Yazdani}}]{Gomes2007}%
  \BibitemOpen
  \bibfield  {author} {\bibinfo {author} {\bibfnamefont {K.~K.}\ \bibnamefont
  {Gomes}}, \bibinfo {author} {\bibfnamefont {A.~N.}\ \bibnamefont
  {Pasupathy}}, \bibinfo {author} {\bibfnamefont {A.}~\bibnamefont {Pushp}},
  \bibinfo {author} {\bibfnamefont {S.}~\bibnamefont {Ono}}, \bibinfo {author}
  {\bibfnamefont {Y.}~\bibnamefont {Ando}}, \ and\ \bibinfo {author}
  {\bibfnamefont {A.}~\bibnamefont {Yazdani}},\ }\href@noop {} {\bibfield
  {journal} {\bibinfo  {journal} {Nature}\ }\textbf {\bibinfo {volume} {447}},\
  \bibinfo {pages} {569} (\bibinfo {year} {2007})}\BibitemShut {NoStop}%
\bibitem [{\citenamefont {Kohsaka}\ \emph {et~al.}(2008)\citenamefont
  {Kohsaka}, \citenamefont {Taylor}, \citenamefont {Wahl}, \citenamefont
  {Schmidt}, \citenamefont {Lee}, \citenamefont {Fujita}, \citenamefont
  {Alldredge}, \citenamefont {McElroy}, \citenamefont {Lee}, \citenamefont
  {Eisaki}, \citenamefont {Uchida}, \citenamefont {Lee},\ and\ \citenamefont
  {Davis}}]{Kohsaka2008}%
  \BibitemOpen
  \bibfield  {author} {\bibinfo {author} {\bibfnamefont {Y.}~\bibnamefont
  {Kohsaka}}, \bibinfo {author} {\bibfnamefont {C.}~\bibnamefont {Taylor}},
  \bibinfo {author} {\bibnamefont {Wahl}}, \bibinfo {author} {\bibfnamefont
  {A.}~\bibnamefont {Schmidt}}, \bibinfo {author} {\bibfnamefont
  {J.}~\bibnamefont {Lee}}, \bibinfo {author} {\bibfnamefont {K.}~\bibnamefont
  {Fujita}}, \bibinfo {author} {\bibfnamefont {J.~W.}\ \bibnamefont
  {Alldredge}}, \bibinfo {author} {\bibfnamefont {K.}~\bibnamefont {McElroy}},
  \bibinfo {author} {\bibfnamefont {J.}~\bibnamefont {Lee}}, \bibinfo {author}
  {\bibfnamefont {H.}~\bibnamefont {Eisaki}}, \bibinfo {author} {\bibnamefont
  {Uchida}}, \bibinfo {author} {\bibfnamefont {D.-H.}\ \bibnamefont {Lee}}, \
  and\ \bibinfo {author} {\bibfnamefont {J.~C.}\ \bibnamefont {Davis}},\
  }\href@noop {} {\bibfield  {journal} {\bibinfo  {journal} {Nature}\ }\textbf
  {\bibinfo {volume} {454}},\ \bibinfo {pages} {1072} (\bibinfo {year}
  {2008})}\BibitemShut {NoStop}%
\bibitem [{\citenamefont {Ghiringhelli}\ \emph {et~al.}(2012)\citenamefont
  {Ghiringhelli}, \citenamefont {Tacon}, \citenamefont {Minola}, \citenamefont
  {Blanco-Canosa}, \citenamefont {Mazzoli}, \citenamefont {Brookes},
  \citenamefont {Luca}, \citenamefont {Frano}, \citenamefont {Hawthorn},
  \citenamefont {He}, \citenamefont {Loew}, \citenamefont {Sala}, \citenamefont
  {Peets}, \citenamefont {Salluzzo}, \citenamefont {Schierle}, \citenamefont
  {Sutarto}, \citenamefont {Sawatzky}, \citenamefont {Weschke}, \citenamefont
  {Keimer},\ and\ \citenamefont {Braicovich}}]{Ghir2012}%
  \BibitemOpen
  \bibfield  {author} {\bibinfo {author} {\bibfnamefont {G.}~\bibnamefont
  {Ghiringhelli}}, \bibinfo {author} {\bibfnamefont {M.~L.}\ \bibnamefont
  {Tacon}}, \bibinfo {author} {\bibfnamefont {M.}~\bibnamefont {Minola}},
  \bibinfo {author} {\bibfnamefont {S.}~\bibnamefont {Blanco-Canosa}}, \bibinfo
  {author} {\bibfnamefont {C.}~\bibnamefont {Mazzoli}}, \bibinfo {author}
  {\bibfnamefont {N.~B.}\ \bibnamefont {Brookes}}, \bibinfo {author}
  {\bibfnamefont {G.~M.~D.}\ \bibnamefont {Luca}}, \bibinfo {author}
  {\bibfnamefont {A.}~\bibnamefont {Frano}}, \bibinfo {author} {\bibfnamefont
  {D.~G.}\ \bibnamefont {Hawthorn}}, \bibinfo {author} {\bibfnamefont
  {F.}~\bibnamefont {He}}, \bibinfo {author} {\bibfnamefont {T.}~\bibnamefont
  {Loew}}, \bibinfo {author} {\bibfnamefont {M.~M.}\ \bibnamefont {Sala}},
  \bibinfo {author} {\bibfnamefont {D.~C.}\ \bibnamefont {Peets}}, \bibinfo
  {author} {\bibfnamefont {M.}~\bibnamefont {Salluzzo}}, \bibinfo {author}
  {\bibfnamefont {E.}~\bibnamefont {Schierle}}, \bibinfo {author}
  {\bibfnamefont {R.}~\bibnamefont {Sutarto}}, \bibinfo {author} {\bibfnamefont
  {G.~A.}\ \bibnamefont {Sawatzky}}, \bibinfo {author} {\bibfnamefont
  {E.}~\bibnamefont {Weschke}}, \bibinfo {author} {\bibfnamefont
  {B.}~\bibnamefont {Keimer}}, \ and\ \bibinfo {author} {\bibfnamefont
  {L.}~\bibnamefont {Braicovich}},\ }\href@noop {} {\bibfield  {journal}
  {\bibinfo  {journal} {Science}\ }\textbf {\bibinfo {volume} {337}},\ \bibinfo
  {pages} {821} (\bibinfo {year} {2012})}\BibitemShut {NoStop}%
\bibitem [{\citenamefont {Torchinsky}\ \emph {et~al.}(2013)\citenamefont
  {Torchinsky}, \citenamefont {Mahmood}, \citenamefont {Bollinger},
  \citenamefont {Bozovic},\ and\ \citenamefont {Gedik}}]{Torchinsky2013}%
  \BibitemOpen
  \bibfield  {author} {\bibinfo {author} {\bibfnamefont {D.~H.}\ \bibnamefont
  {Torchinsky}}, \bibinfo {author} {\bibfnamefont {F.}~\bibnamefont {Mahmood}},
  \bibinfo {author} {\bibfnamefont {A.~T.}\ \bibnamefont {Bollinger}}, \bibinfo
  {author} {\bibfnamefont {I.}~\bibnamefont {Bozovic}}, \ and\ \bibinfo
  {author} {\bibfnamefont {N.}~\bibnamefont {Gedik}},\ }\href@noop {}
  {\bibfield  {journal} {\bibinfo  {journal} {Nature Materials}\ }\textbf
  {\bibinfo {volume} {12}},\ \bibinfo {pages} {387} (\bibinfo {year}
  {2013})}\BibitemShut {NoStop}%
\bibitem [{\citenamefont {Nikoli{\'c}}\ \emph {et~al.}(2013)\citenamefont
  {Nikoli{\'c}}, \citenamefont {Duri{\'c}},\ and\ \citenamefont
  {Tesanovi{\'c}}}]{Nikolic2011a}%
  \BibitemOpen
  \bibfield  {author} {\bibinfo {author} {\bibfnamefont {P.}~\bibnamefont
  {Nikoli{\'c}}}, \bibinfo {author} {\bibfnamefont {T.}~\bibnamefont
  {Duri{\'c}}}, \ and\ \bibinfo {author} {\bibfnamefont {Z.}~\bibnamefont
  {Tesanovi{\'c}}},\ }\href@noop {} {\bibfield  {journal} {\bibinfo  {journal}
  {Physical Review Letters}\ }\textbf {\bibinfo {volume} {110}},\ \bibinfo
  {pages} {176804} (\bibinfo {year} {2013})}\BibitemShut {NoStop}%
\bibitem [{\citenamefont {Nikoli{\'c}}(2014)}]{Nikolic2014}%
  \BibitemOpen
  \bibfield  {author} {\bibinfo {author} {\bibfnamefont {P.}~\bibnamefont
  {Nikoli{\'c}}},\ }\href@noop {} {\bibfield  {journal} {\bibinfo  {journal}
  {Physical Review A}\ }\textbf {\bibinfo {volume} {90}},\ \bibinfo {pages}
  {023623} (\bibinfo {year} {2014})}\BibitemShut {NoStop}%
\bibitem [{\citenamefont {Nikoli{\'c}}(2016)}]{Nikolic2014a}%
  \BibitemOpen
  \bibfield  {author} {\bibinfo {author} {\bibfnamefont {P.}~\bibnamefont
  {Nikoli{\'c}}},\ }\href@noop {} {\bibfield  {journal} {\bibinfo  {journal}
  {Physical Review B}\ }\textbf {\bibinfo {volume} {94}},\ \bibinfo {pages}
  {064523} (\bibinfo {year} {2016})}\BibitemShut {NoStop}%
\bibitem [{\citenamefont {Nikoli{\'c}}(2020{\natexlab{b}})}]{Nikolic2019b}%
  \BibitemOpen
  \bibfield  {author} {\bibinfo {author} {\bibfnamefont {P.}~\bibnamefont
  {Nikoli{\'c}}},\ }\href@noop {} {\bibfield  {journal} {\bibinfo  {journal}
  {Physical Review B}\ }\textbf {\bibinfo {volume} {102}},\ \bibinfo {pages}
  {075131} (\bibinfo {year} {2020}{\natexlab{b}})}\BibitemShut {NoStop}%
\bibitem [{\citenamefont {Zinn-Justin}(2005)}]{ZinnJustin2001}%
  \BibitemOpen
  \bibfield  {author} {\bibinfo {author} {\bibfnamefont {J.}~\bibnamefont
  {Zinn-Justin}},\ }\href@noop {} {\bibfield  {journal} {\bibinfo  {journal}
  {Lecture Notes in Physics}\ }\textbf {\bibinfo {volume} {659}},\ \bibinfo
  {pages} {167} (\bibinfo {year} {2005})},\ \bibinfo {note} {topology and
  Geometry in Physics, Ed. Eike Bick and Frank Daniel Steffen}\BibitemShut
  {NoStop}%
\end{thebibliography}

%

\end{document}